\documentclass[aps,prd,showpacs,twocolumn,nofootinbib,superscriptaddress]{revtex4-2}
\usepackage{amsmath,amssymb,amsfonts}
\usepackage{bm}
\usepackage{verbatim}
\usepackage{hyperref}
\usepackage{slashed,braket}
\usepackage{ulem}
\usepackage{cancel}
\usepackage{color}
\usepackage{graphicx,graphics}
\usepackage[title,titletoc]{appendix}
\usepackage{stackengine}
\usepackage{mathrsfs}
\usepackage{tikz-feynman}
\usepackage{tikz}
\usepackage{pgfplots}

\newcommand{\tcH}{\tilde{\mathscr{H}}}
\newcommand{\tE}{\tilde{E}}

\newcommand{\tr}{\operatorname{tr}}
\newcommand{\sign}{\operatorname{sign}}

\renewcommand{\Re}{\operatorname{Re}}

\newcommand{\ep}{\epsilon}

\newcommand{\w}{\omega}
\newcommand{\D}{\Delta}
\newcommand{\al}{\alpha}

\newcommand{\bk}{\mathbf{k}}

\newcommand{\hl}[1]{{#1}}

\newcommand{\added}[1]{{#1}} 
\newcommand{\fin}[1]{{#1}}

\begin{document}

\title{Magnetoconductivity of Dirac semimetals and chiral magnetic effect from Keldysh technique}

\author{Ruslan A.~Abramchuk}
\email{abramchukrusl@gmail.com}
\affiliation{Physics Department, Ariel University, Ariel 40700, Israel}

\author{M.A.Zubkov}
\email{mikhailzu@ariel.ac.il}
\affiliation{Physics Department, Ariel University, Ariel 40700, Israel}

\date{\today}

\begin{abstract}
    Negative magnetoresistance in Dirac semimetals is typically considered as a manifestation of chiral magnetic effect (CME). 
    The relation between these two phenomena has the status of a hypothesis and is based on  sequence of assumptions.
    We rely on the Keldysh technique of non-equilibrium theory. 
    It allows us to investigate the accumulation of axial charge 
        ---the process that involves chiral anomaly and relaxation followed by the energy dissipation. 
    We consider the case of strong magnetic field and calculate directly axial charge density and electric conductivity,
        taking into account scattering on impurities and interaction with phonons. 

\hl{
    Overall, our analysis is consistent with the CME-based analysis of magnetoconductivity in Dirac semimetals in the limit of strong magnetic fields.
    We obtain the same relation 
        between axial charge density and electric and magnetic fields, 
        as well as between electric current and axial charge density 
        as the standard heuristic CME calculation. 
    But we also calculate the relaxation time as a function of model parameters,
        instead of introducing the relaxation time as a free parameter.
    }
\end{abstract}

\maketitle

\section{Introduction}
\label{SectIntro}

Negative magnetoresistance is one of the characteristic features of Weyl {and} Dirac semimetals. According to the common lore it is related to the chiral magnetic effect. The derivation {\cite{CMEZrTe5,Kharzeev2017}} of relation between these two phenomena is associated with the following basic assumptions:

\begin{enumerate}

\item

In the presence of parallel electric and magnetic fields  $\vec E$, $\vec B$ the chiral anomaly pumps pairs (left-handed electron and right-handed hole or vice versa) from the Dirac sea of occupied energy levels. Qualitatively it is clear that this process results in the appearance of chiral imbalance (i.e. the difference between the densities $\rho_R$, $\rho_L$ of left-handed and right-handed fermion excitations). Naive calculations result in the following expression for the density of chiral charge:
\begin{equation}
    \rho_5 = \rho_R - \rho_L = N_f\frac{e^2{\vec E} \cdot{\vec B}}{{2\pi^2}\hbar^2}\,\, \tau_5, \label{EqRho5CME}
\end{equation}
        where $N_f$ is the number of Dirac fermions in the given system, while $\tau_5$ is a relaxation time. 
{\it In this paper we use SI units.}

\item 

The next general assumption is that the chiral imbalance is related to the notion of chiral chemical potential $\mu_5 = (\mu_R - \mu_L)/2$ in the way similar to relation of the total number of electrons and the ordinary chemical potential. Namely, assuming the presence of the corresponding steady state we have {in case of weak magnetic field limit}
\begin{equation}
    \rho_5 \approx \frac{\mu_5}{3 v_F^3\hbar^3}\Big(T^2 + \frac{\mu^2}{{\pi^2}} \Big), \label{EqMu5CMEw}
\end{equation}
where $v_F$ is Fermi velocity. Notice that in the present paper we express temperature in the units of energy, i.e. we assume that $T = k_B T^o$, where $T^o$ is expressed in K.
In the  strong magnetic field limit, 
    {if} only the lowest Landau level contributes to the dynamics { \cite{Kharzeev2017} }
\begin{equation}
	\rho_5 \approx \added{N_f}\frac{\mu_5}{2 \pi^2 {v_F\hbar^2}} |e{\vec B}|. \label{EqMu5CME}
\end{equation}

 \item
 
 Next, the standard expression for the electric current of the CME is used 
 
 \begin{equation}
 {\vec j} = \added{N_f}\frac{e^2}{2\pi^2\hbar^2} \mu_5 {\vec B}\label{CME}
 \end{equation}
 
{In the limit of strong magnetic field}  the following dependence of electric conductivity on magnetic field {is obtained} $\sigma^{ij}_{CME} \sim B^i B^j/|B|$. 
For the component along the magnetic field
{\begin{equation}
    \sigma^{zz}_{CME} =  N_f \frac{e^3 v_F|{\vec B}|}{{2\pi^2}\hbar^2}  \tau_5 \label{EqCond2CME}
\end{equation}}
which was also obtained from Kubo formula in \cite{gorbar2014chiral},\hl{\cite{Lu_2015,Li_2023}} as the magnetoconductivity of Weyl fermion system, without any reference to the CME.
       
\end{enumerate}   

There exist many versions of the calculations that relate negative magneto-resistance in Weyl and Dirac semimetals {and vector currents in QCD to the chiral magnetic effect \cite{Fukushima2008,Kharzeev2017,Stephanov2012ki,Burkov2014,Son2013,Gorbar2016,Gao2012,Lin2019,Hattori2016lqx,Huang2017}. 
In particular, the so-called chiral kinetic theory was used to obtain the similar result \cite{Son_2012,Stephanov2012ki,Gao2012,PhysRevB.96.235134,sekine2021axion} 
or analyze underlying quantum dynamics \cite{Hattori2016lqx,Lin2019},
as well as quasiclassical approaches \cite{Burkov2014,Son2013,Gorbar2016} }
To the best of our knowledge all these calculations are based on various versions of  the three mentioned above  assumptions, which, in turn, have the status of hypotheses. {It is worth mentioning that in $^3$He-A superfluid (where the quasiparticles are electrically neutral) there exists a special type of chiral magnetic effect due to emergent gauge field \cite{volovik2017chiral}.} 

Eq. (\ref{EqRho5CME}) is obtained as the solution of a phenomenological kinetic equation that accounts for the  chirality relaxation. This equation should appear as a certain approximation to the kinetic equation derived from (non-equilibrium) quantum field theory,
    that can be addressed with {the} Keldysh--Schwinger technique
    (the Non-equilibrium Diagram Technique---NDT). 

In the original derivation of Eq. (\ref{EqMu5CME}) the naive continuum theory of non-interacting massless Dirac fermions is used. It assumes that the left-handed and the right-handed fermions exist separately in thermal equilibrium. 
Strictly speaking, this supposition is not correct due to the chiral anomaly. Therefore, Eq. (\ref{EqMu5CME}) might express the result for a semi-equilibrium steady state, rather than for the true equilibrium. The very notion of chiral chemical potential in this setup is not well-defined.  

Eq. (\ref{CME}) originally was proposed as {an} expression for the electric current in the presence of external magnetic field and chiral chemical potential. It has been proven later using several different methods {\cite{Valgushev:2015pjn,Buividovich:2015ara,Buividovich:2014dha,Buividovich:2013hza,Z2016_1,Z2016_2,nogo,nogo2,BLZ2021}} that in {the} true equilibrium the CME is absent, and Eq. (\ref{CME}) is not valid (see also recent results of numerical lattice simulations \cite{brandt2024absence,buividovich2024out}). However, out of equilibrium the CME is { present} \cite{BLZ2022}. For example, for the system with time depending chiral chemical potential the electric current is nonzero. For the oscillating $\mu_5$ with frequency $\omega$ Eq. (\ref{CME}) is recovered in the limit $\omega \to 0$ if the limit of infinitely large system volume is taken first. Therefore, it is possible that Eq. (\ref{CME}) remains alive in {an}other non-equilibrium context of the steady state that exists in the presence of external electric field, when energy produced by it is dissipated during the annihilation of the electrons and holes of opposite chirality. 

We consider continuum Dirac fermion as the low energy effective theory for charge carriers in a vicinity of a Fermi point in Dirac semimetals. 
Chiral anomaly applies to such degrees of freedom.

As we show in this paper, 
    the mechanism for the annihilation of electrons and holes of opposite chirality is defined by 
    the region of Brillouin zone situated far from the Fermi points, 
    where the chiral symmetry is broken naturally. 
Surprisingly, the most obvious mechanism of chirality breaking---the Dirac mass,
    happens to be unnecessary in order to have finite results.
We retain the Dirac mass, which introduces the band gap at the Fermi point,
    but assume the mass to be small in comparison to the chemical potential
    (\(\mu = E_F-E_\Gamma\) the distance between the Fermi level and the Fermi point position)
\begin{equation}
    mv_F^2 \ll |\mu|. \label{EqApprMum}
\end{equation}
In Dirac semimetals, e.g.~in \(ZrTe_5\) \cite{CMEZrTe5,zhang2021observation}, \(Cd_3As_2\) \cite{Crassee2018}, \(\mathrm{Mn}_{1-x} \mathrm{Ge}_x \mathrm{Bi}_2 \mathrm{Te}_4\) \cite{shikin2024}  the gap  is of order of meV, 
    which {may be} smaller than the room temperature (\(\sim 25\) meV).
The chemical potential and its exact value is a property of a given crystal.
In \(ZrTe_5\) in which the magneto-conductivity was {originally} observed, the chemical potential is \(\mu\approx 100\) meV \cite{CMEZrTe5} and the Fermi velocity \(v_f \sim c/300\).

Magnetoconductivity, or negative magneto-resistance (NMR), was also observed in Weyl semimetal NbP \cite{niemann2017chiral}, 
    in which the Fermi point is not so deep inside the Fermi sphere \(\mu \sim 10 \text{ meV} \sim 100\) K,
    as well as in Dirac semimetals \(Cd_3As_2\) \cite{li2016negative} and \(Na_3Bi\) \cite{xiong2015evidence},
    in which the Fermi level was purposefully brought down to the Fermi point.
The effect in those cases is saturated by temperature rather than chemical potential 
    (the effect is thermal-activated in those examples),
    and thus is not addressed in this paper.
In case of thermal-activated effect NMR may even grow \cite{niemann2017chiral} with temperature, 
    while in the system \(ZrTe_5\) \cite{CMEZrTe5},
    which is mainly considered in this paper,
    magnetoconductivity diminishes with temperature.

We perform calculations in the limit of strong magnetic field,
    as compared to the scale of chemical potential
\begin{equation}
    {|\mu|} \lesssim v_F \sqrt{2| e B| \hbar},   \label{EqApprMuB}
\end{equation}
that is $l_B \lesssim \sqrt{2} l_\mu$. 
In general, several lengths are to be compared, including  
{\it the mean free path length $l_0$, the magnetic length $l_B = \sqrt{\frac{\hbar}{|e B|}}$, the thermal length $l_T = \frac{\hbar v_F}{k_B T} $, and $l_\mu = \frac{\hbar v_F}{\mu}$.} 
Provided that also $l_B \lesssim l_T$ in this case only the lowest Landau level contributes  the physical observables.

For example, for Cd$_3$As$_2$ Fermi velocity $v_F \sim c/200$, the mean free path may reach $l_0 \sim 100 \mu$m, and at $T = 10 $K, $\mu \sim 100$ meV and $B = 1 $T we have $l_B \sim 6 \times 10^{-8}m \sim l_\mu \sim 6\times 10^{-8}m \ll 10^{-5}m = l_T \ll l_0 \sim 10^{-4}m $, which means that {the} magnetic scale and the scale of the chemical potential are of the same orders of magnitude, and are much larger than the scales of mean free path and temperature. 
Throughout the text we use the example of $Cd_3As_2$ as a reference in order to evaluate the order of magnitude of the obtained physical values. 
We will also use for the same purpose the case of \(ZrTe_5\), in which the magnetoconductivity was observed. 
For other Weyl/Dirac semimetals such estimates may be accepted  as rough estimates.  

The dissipation of energy is parametrized with the imaginary part of electron self-energy,
    which is the electron energy level half-width. 
We show how the imaginary part appears due to interaction with an ensemble of phonons and scattering on impurities. 
The corresponding calculation substantially differs from the case of QED \cite{Shovkovy2024},
    since a characteristic feature of semimetals is relatively small speed of sound (velocity of phonons) in comparison to the Fermi velocity (of electrons). 

Though the expression for strong field magnetoconductivity Eq.\eqref{EqCond2CME} was obtained in various ways, 
    the expression alone fails to match any experimental data on the dependence of conductivity on magnetic field and temperature in any parameters range,
    since parameter $\tau_5$ itself depends on magnetic field and on temperature. 
In the present paper we confirm Eq.\eqref{EqCond2CME}, 
    and calculate dependence of parameter \(\tau_5\) on temperature and on magnetic field 
    in a model that includes disorder (weak disorder of delta-potential non-magnetic impurities) and phonons (thermal ensemble of longitudinal acoustic phonons). 
The obtained result appears to match the experimental data in a certain parameters range, 
    as discussed in Section \ref{SectCon}.

\hl{Strong field magnetoconductivity of Weyl and Dirac semimetals was addressed with Kubo formula for various kinds of disorder in \cite{Li_2023}.
For the zero scattering length uncorrelated disorder model, which we use in this paper, the same dependence on magnetic field was found.  
Additionally, possible dependence of the Fermi velocity itself on magnetic field was analyzed in \cite{Lu_2015}, 
    which is beyond the scope of this paper.}

Overall, we believe that the three assumptions listed at the beginning of this section describe the real situation in solids qualitatively correct. 
The lack of rigorous derivation of this pattern, however, leaves a gap in understanding of magnetoconductivity
    (e.g.~the phenomenological relaxation time). 
In the present paper we are trying to close the gap and calculate the strong field magnetoconductivity of Dirac semimetals using the rigorous Keldysh-Schwinger technique,
    which is free from the aforementioned uncertainties of the phenomenological approach. 
We also suppose that our calculation remains valid for the wider class of type-I Weyl semimetals. 

The paper is organized as follows.
In the next Section we briefly introduce the Keldysh technique notation for Dirac fermions, 
    and some other mathematical tools and notation.
In Section \ref{SectCond} we derive formulae for electric conductivity and axial charge density in strong magnetic field limit as functions of the level width,
    and also extensively discuss the chirality relaxation mechanism in Section \ref{SectCondZG}.
In Section \ref{SectSelf} we calculate the level width from interaction with impurities and phonons.
In Section \ref{SectCon} we sum up the results, compare them to experimental data, discuss limitations and future prospects.

\section{The Keldysh technique}\label{SectNDT}

In this section we introduce the Keldysh formalism for Dirac fermions in the ``upper-triangle'' representation as in \cite{Arseev2015,Banerjee2020obs},
    and generally follow \cite{Kamenev2005course,KamenevBook}
Then we consider the strong magnetic field limit.

Starting with the retarded Green function for a free theory (in our case, for free Dirac fermions),
    other components of the Keldysh propagator \(\hat G\) can be written down,
    and its inverse Keldysh operator \(\hat Q\) is introduced
\begin{gather}
    \hat G = \begin{pmatrix} G^R & G^< \\ 0 & G^A \end{pmatrix}
        = \hbar\hat Q^{-1}, 
        \quad \hat Q = \begin{pmatrix} Q^R & Q^< \\ 0 & Q^A \end{pmatrix},
        \label{EqGss}\\
    {\hbar}\begin{pmatrix} 1 &  \\   & 1 \end{pmatrix} 
        = \begin{pmatrix} Q^R & Q^< \\   & Q^A \end{pmatrix}
          \begin{pmatrix} G^R & G^< \\   & G^A \end{pmatrix} . \label{EqGAR} \\
    G^{R(A)} = \hbar Q^{R(A)-1}, \quad {G^< = -{\hbar^{-1}}G^RQ^<G^A}.
\end{gather}
Such definition, suggested in \cite{Kamenev2005course}, is convenient for the path-integral representation of the Keldysh formalism.
The retarded (advanced) components trivially follow from the free Dirac propagator
\begin{gather}
    Q^{R(A)} = p_0 \pm i\epsilon - (v_F\vec{\alpha}\cdot\vec p + \gamma_0 m v_F^2),\nonumber \\
\end{gather}
where the matrices \(\vec{\alpha}=\gamma_0\vec{\gamma}\), 
    and \(\gamma_\mu\) are the gamma matrices with the Minkowski signature.
$\epsilon \to +0$ ensures proper analytic properties of the Green functions.

The `lesser' Keldysh--Green function can be obtained easily from the rule 
    that the lesser component of the free Keldysh operator is pure regularization \cite{Kamenev2005course}
\begin{gather}
	Q^< = -2i\epsilon n(p_0).
\end{gather}
With  the initial spin-independent isotropic distribution $n(p_0)$ 
    the standard result emerges 
    (more complicated situation can be also addressed using this rule)
\begin{align}
    G^< &= (G^A-G^R)n(p_0) \label{EqGLAR}\\
    &\approx 2\pi \hbar i\Delta_\epsilon{(p_0, E)} \nonumber \\
    &\quad \times(p_0+v_F\vec{\alpha}\cdot \vec p + \gamma_0 {v_F^2}m)n(p_0),
    \label{EqGless}
\end{align}
where \(E=\sqrt{v_F^2\vec p^2+v_F^4m^2}\) is the spectra.
The equilibrium distribution function at finite temperature is given by the Fermi distribution 
\begin{equation}
	n(p_0)= (e^\frac{p_0-\mu}{T}+1)^{-1} \to \Theta(\mu - p_0)  \label{EqFermi}.
\end{equation}  
In this paper, the limit \(|\mu|\gg T\) is relevant, and the distribution is degenerate.

The broadened delta-function
\begin{gather}
    \Delta_\ep{(p_0,E)} = \frac{\delta_\ep(p_0-E) - \delta_\ep(p_0+E)}{2E},\label{EqDeltaEps}\\
    \delta_\epsilon(x)=\frac{\epsilon}{\pi(x^2+\epsilon^2)},
\end{gather}
produces the delta-function at vanishing \(\ep\to +0\),
\(\delta_\epsilon(x)\to \delta(x)\),
\(\Delta_\ep{(p_0,E)}\to \sign(p_0)\delta(p_0^2-E^2)\).

The approximate expression of Eq. (\ref{EqGless}) will be used with {the} bare dissipation rate $\epsilon \to 0$ replaced by the finite one extracted from the fermion self-energy, 
    which is assumed to be much smaller than $\mu$.  
The finite level half-width, or dissipation rate, $\epsilon$ appears through the imaginary part of self-energy (e.g.~as is explained in the general context in \S 123 of \cite{Landau2013statistical}).

To calculate analytically the `lesser' component of a product of Keldysh--Green functions,
we employ the formula \cite{Arseev2015,Banerjee2020obs}
\begin{equation}
    \big(\prod_{i=1}^n\hat K_i\big)^< 
        = \sum_{l=1}^n\big(\prod_{i=1}^{l-1}K_i^R\big) 
        ~ K_l^< \big(\prod_{j=l+1}^{n}K_j^A\big).
\end{equation}
{In this expression the product  $\big(\prod_{i=1}^{l-1}K_i^R\big)$ for $l<2$ is assumed to be equal to unity.} For example, for $n = 2$ the formula reads     
\((\hat K_1\hat K_2)^< = K_1^RK_2^< + K_1^<K_2^A\).

If $\hat{G}$ or $\hat{Q}$  is written without arguments, it is said of an operator, while the Green function itself and its inverse are the matrix elements: $\hat{G}(x,y) = \langle x | \hat{G} | y\rangle$, $\hat{Q}(x,y) = \langle x | \hat{Q} |y \rangle $ (or the Fourier transforms $\hat{G}(p,q) = \langle p | \hat{G} | q\rangle$ and $\hat{Q}(p,q) = \langle p | \hat{Q} | q\rangle$).  
Correspondingly, in Eq. (\ref{EqGAR}) the products of the mentioned operators are assumed. 
If the Green function is translation-invariant, in momentum space we have $\hat{G}(p,q) = \hat{G}'(p-q)\delta(p-q)$. 
We will omit the prime symbol ($'$), and denote the Green function itself and $\hat{G}'$ by the same letter.

In this paper we also use the Wigner-Weyl calculus \hl{(see, for example, \cite{chernodub2017scale,zhang2020influence})}.
The Wigner transformation of matrix elements of an operator $A$ is defined as 
\begin{equation}
A_W(x|p)=\int d^{D+1} y\, e^{i y^\mu p_\mu/\hbar }\langle x+y/2|{A}|x-y/2ngle
\label{WignerTr}
\end{equation}
Wigner transform of product of operators results into the star (Moyal) product
\begin{align}
    ( A B)_W&(x|p) = A_W(x|p)\star B_W(x|p) \label{EqStarProd}\\
    &= A_W(x|p)\,e^{{-}i(\overleftarrow{\partial}_{x^{\mu}}\overrightarrow{\partial}_{p_{\mu}}-\overleftarrow{\partial}_{p_{\mu}}\overrightarrow{\partial}_{x^{\mu}})/2\hbar}B_W(x|p).
        \nonumber
\end{align}

In strong magnetic field limit Eq.\eqref{EqApprMuB}, at small enough temperatures,
    all the dynamics occurs on the lowest Landau level (LLL).
In this case, the system is in some sense effectively 1+1 dimensional.
\hl{The corresponding approximate propagator can be constructed, for example, 
    by definition, from the plane waves solution of the Dirac equation in constant uniform magnetic field.
However, instead we use the results and formalism developed in \cite{Gusynin1999pq}.
}

In the symmetric gauge $\vec{A}(\vec{r}) = \frac{1}{2}~\vec{r} \times \vec{B} $ the Feynman LLL propagator for a Dirac fermion in magnetic field reads 
    (see Eq. (16) in \cite{Gusynin1999pq}; \(B=|\vec B|=|B_z|\), we assume that magnetic field is directed along the third axis)
\begin{eqnarray}
	G_F(x,y) & = & e^{\frac{i}{\hbar}\int_x^y dz eA(z)} S(x-y)\nonumber\\&=&{e^{-\frac{i}{\hbar}(x-y) eA((x+y)/2)} S(x-y) },
\end{eqnarray}
where the exponential represents  parallel transporter along the straight line connecting the end-points $x$ and $y$, 
    and the Fourier transformation of $S$ reads
\begin{equation}
    -iS_\text{LLL}(p) = 2e^{-\frac{p_\bot^2}{{e \hbar}B}}
        \frac{p_0\gamma_0-{v_F}p_3\gamma_3{+mv_F^2}}{p_0^2-{v_F^2}p_3^2{-m^2v_F^4}+i0}O^{-},
\end{equation}
where the Dirac algebra matrix 
\begin{equation}
    O^{-} = \frac12(1 - i\gamma_1\gamma_2\sign(e B_3))  \label{EqHProjector}
\end{equation}
is the projection operator on the fermion states with the spin polarized along the magnetic field.
The LLL fermions couple only to the components of
    the electric field parallel to the magnetic field \cite{Gusynin1999pq}
    because of the projection operator \(O^-\) 
\begin{equation}
    O^-\gamma^\mu O^- = O^-\gamma^\mu_\parallel.
\end{equation} 

Then, we write the corresponding Keldysh-Green functions in the LLL approximation
for the electrons in a Dirac semimetal (in the absence of electric field):
\begin{eqnarray}
	\hat{G}(x,y) &=& e^{\frac{i}{\hbar}\int_x^y dz eA(z)} \hat{G}^{(LLL)}(x-y)\nonumber\\  &=& {e^{-\frac{i}{\hbar}(x-y) eA((x+y)/2)} \hat{G}^{(LLL)}(x-y)}.\label{GRAL0}
\end{eqnarray}
For brevity we will omit in the following the subscript $(LLL)$ unless it causes confusion. Then the  translational invariant parts of the Green functions will be denoted by the same letters as the complete ones. Their Fourier transforms (placed in the $2\times 2$ Keldysh matrix) read:
\begin{align}
			\hat G^\text{(LLL)}(p_0,p_3,p_\bot)
		\approx 2 e^{-\frac{p_\bot^2}{|{e\hbar}B|} } {\tilde G}({p_0,p_3})O^-.
		\label{EqSLLL}
\end{align}
We denote 
\begin{equation}
	G_\bot(p_\bot) = 2 e^{-\frac{p_\bot^2}{|{e\hbar}B|} }
\end{equation}
In the strong magnetic field the notation for the reduced 2D Keldysh-Green functions is used
\begin{gather}
    \w_\pm=p_0\pm i\ep, \quad \w=p_0, \quad \tilde G^{R(A)} = \hbar(\w_\pm -\tcH)^{-1}, \nonumber\\
    \tcH = v_F\al_3p_3 + \gamma_0 m v_F^2, \quad
        \tE=\sqrt{v_F^2p_3^2+m^2}, \label{wH1}\\
    \D_{\ep}=\D_{\ep}(\w,\tE),\quad \tilde G^< = 2\pi \hbar i{(\w+\tcH)\Delta_{\ep}} n(\w). \nonumber
\end{gather}

{Due to Eq. (\ref{GRAL0}) we obtain the following expression for the Wigner transformation of Green function:
	\begin{eqnarray}
		\hat{G}_W(x|p) = \hat{G}^{(LLL)}(p - e A(x))
	\end{eqnarray}
The Weyl symbol of its inverse (i.e. Wigner transformation of the matrix elements of $\hat{Q}$) exists,
    however we never use it explicitly.

Let us denote with bold symbols Keldysh functions for interacting theory, up to some order of perturbation theory. 
In Sections \ref{SectEpsH} and \ref{SectImp} we show that the interactions preserve the general form of the Green function, and
\begin{eqnarray}
	\hat{\bf G}(x,y) &=& e^{\frac{i}{\hbar}\int_x^y dz eA(z)} \hat{\bf G}^{(LLL)}(x-y).\label{GRAL}
\end{eqnarray} 
The Fourier transformation of $\hat{\bf G}^{(LLL)}$ is given by
\begin{align}
	\hat {\bf G}^\text{(LLL)}(p_0,p_3,p_\bot)
    \approx G_\bot(p_\bot) \tilde{{{\bf G}}}_{p_0p_3} O^-. \label{EqGLLL}
\end{align}
while 
\begin{equation}
\tilde{{\bf {G}}}^{-1}(p_0,p_3) = {\tilde{ G}}^{-1}(p_0,p_3) -  \hat{\Sigma}(p_0)	
\end{equation}
We show in Appendix \ref{SectAppC} that $\hat{\Sigma}(p_0)$ contains only unity matrix and $\gamma^0$. 

By \(\tilde{\bf G}\) we denote the $2\times 2$ matrix in Keldysh space, 
    and omit the extra hat symbol in $\tilde{\bf G}$ for brevity. 
\begin{equation}
\tilde{\bf G} = \left( \begin{array}{cc}\tilde{\bf G}^R & \tilde{\bf G}^< \\ 0 & \tilde{\bf G}^A\end{array} \right)
\end{equation}
At least in the first order of perturbation theory, just like in the absence of interactions
 \begin{equation}
 	{\tilde { {\bf G}}}^< = ({\tilde { {\bf G}}}^A-{\tilde { {\bf G}}}^R) n(p_0).
 \end{equation}

Electrons energy dissipation due to interactions with phonons and disorder, 
    results in broadening of the `lesser' Green function, 
    as explained in Section \ref{SectSelf}.
In particular, for the electrons in the presence of external magnetic field, in the chiral limit $m=0$ on the lowest Landau level 
\begin{align}
	{\bf G}^<(p_0,p_3) 
		&\approx 2\pi \hbar i\Delta_{\epsilon_B}{(p_0- \tilde{E}({ p_3}))} \nonumber \\
	&\quad \times(p_0+v_F{\alpha}_3 p_3)n(p_0),
\end{align} 
where $\tilde{E}(p_3) = v_F^2 |{p}_3|$, and $\Delta_{\epsilon_B}(p_0- \tilde{E}({p}_3))$ is given by Eq.\eqref{EqDeltaEps} with some finite level half-width \(\ep=\ep_B\).

Such energy level broadening results in an important qualitative change of the system. 
For the purpose of illustration let us consider the case of degenerate Fermi distribution Eq. \eqref{EqFermi}. 
In the absence of dissipation $\epsilon_B \to 0$,
    the system represents a collection of non-interacting electrons 
    with all energy levels below $\mu$ occupied, and the remaining levels vacant. 
At a finite $\epsilon_B$ this pattern is broken completely. 
Fourier transformation of the lesser Green function $G^<(p_0,p_3)$ with respect to $p_0$ represents evolution in time of the state with given momentum $p_3$. 
At $\epsilon_B = 0$ the evolution is given by the phase factor $e^{- i\tilde{E}(p) t}$ 
    ---the value of momentum remains the same during evolution in time.
At finite $\epsilon_B$ the evolution contains damping factor 
    that goes to zero at $t \to \infty$  
    with decay time $\frac{\hbar}{2\epsilon_B}$
    ---the probability to find the particle in the state with the same value of momentum goes to zero at $t\gg \frac{\hbar}{2\epsilon_B}$. 
Thus, the filled circles of Fig. \ref{FigBrZ} do not represent the truly occupied states in the presence of finite dissipation rate. 
Rather, the electron states corresponding to these circles have finite lifetime, 
    which means that an electron being placed into such a state disappears from it with finite probability. 
Essentially, an electron cannot disappear completely from the system---
    it jumps to another state, 
    which, in turn, was vacated by another electron for the same reason of a finite lifetime. 
Thus, a filled circle represents a state vacant with a certain probability, 
    and another electron may jump into it, and occupy it (temporary). 
The change of electrons state, labeled with momentum value $p_3$, is accompanied by a phonon emission (or  scattering on impurities) that carries away the extra momentum.

Now we are equipped with the notation to calculate the linear response to electric field,
    for a given level width, 
    which is calculated separately in Section \ref{SectSelf}.

\section{Chiral imbalance and  electric conductivity}\label{SectCond}

In this section we consider the system linear response to constant uniform electric field,
    calculate steady-state axial charge density and electric conductivity.
At the end of this section we extensively discuss the chirality relaxation mechanism.

In the presence of exact chiral symmetry, 
    in external electric and magnetic field, 
    the chiral charge growths with time \(\rho_5(t)\sim E^jB_jt\),
    as can be anticipated from the chiral anomaly. 

If the chiral symmetry is explicitly broken {by the gap parameter \(m\) as well as by regularization},
    and the energy dissipates with {a} rate \(\epsilon_B\) (estimated in Sect. \ref{SectSelf}
        in a strong magnetic field),
we expect saturation of the chiral imbalance.
Thus, we assume that the chiral density and magnetoconductivity 
    can be calculated with the perturbation theory. 

In addition, in lattice regularization the left-handed branch of spectrum is connected to the right-handed branch through the boundary of the Brillouin zone. Therefore, the jumps of electrons on both branches result in the flow of {(almost) occupied} states inverse to the one produced by the chiral anomaly. This is the mechanism of chirality relaxation that occurs even in the limit $m = 0$.  
The corresponding population density on the lowest Landau level (LLL) is depicted in Fig. \ref{FigBrZ}.

With the Keldysh technique, we calculate the response to the electric field 
\begin{align*}
    j_i &=
	\vcenter{\hbox{
		\begin{tikzpicture}
			\begin{feynman}
				\node [empty dot,label=left:\(\gamma_i\)] (p) at (-0.5,0);
				\vertex [label= above:\(\pm\)](v2) at (1,0);
				\node [crossed dot,label=right:\(E_j\)] (e2) at (1.8,0);
				\diagram*{
					(p) --[fermion, very thick, quarter left](v2) --[fermion, very thick, quarter left] (p),
					(e2) --[photon] (v2),
				};
			\end{feynman}
		\end{tikzpicture}
    }}
	\sim E_jB^jB_i 
\end{align*}
Interaction with the thermal bath of phonons (see Section \ref{SectEpsH})
    and scattering on the simplest delta-function non-magnetic impurities (see Section \ref{SectImp})
    are treated perturbatively and embodied into the finite level width \(ep_B\),
while the interaction with magnetic field is treated non-perturbatively, and embodied in the structure of LLL Keldysh propagator Eq.\eqref{EqGLLL}
\begin{align*}
	\vcenter{\hbox{
			\begin{tikzpicture}
				\begin{feynman}
					\vertex [label=below:\(s_1\)](v1) at (0,0);
					\vertex [label=below:\(s_2\)](v2) at (1.1,0);
					\diagram*{
						(v1) --[fermion, very thick](v2),
					};
				\end{feynman}
			\end{tikzpicture}
	}} = 
	\vcenter{\hbox{
			\begin{tikzpicture}
				\begin{feynman}
					\vertex [label=below:\(s_1\)](v1) at (0,0);
					\vertex [label=below:\(s_2\)](v2) at (1.1,0);
					\diagram*{
						(v1) --[fermion](v2),
					};
				\end{feynman}
			\end{tikzpicture}
	}} &+ 
	\vcenter{\hbox{
			\begin{tikzpicture}
				\begin{feynman}
					\vertex [label=below:\(s_1\)](v0) at (-1,0);
					\vertex [label=below:\(s'\)](v1) at (-0.5,0);
					\vertex [label=below:\(s''\)](v2) at (0.5,0);
					\vertex [label=below:\(s_2\)](v3) at (1,0);
					\diagram*{
						(v0) --[fermion](v3),
						(v1) --[scalar, half left] (v2),
					};
				\end{feynman}
			\end{tikzpicture}
	}} +\nonumber\\
	&+ \vcenter{\hbox{
			\begin{tikzpicture}
				\begin{feynman}
					\vertex [label=below:\(s_1\)](v1) at (-1,0);
					\vertex [label=below:\(s_2\)](v2) at (1,0);
					\node [square dot, label=below:\tiny{Imp.}] (n1) at (0,0);
					\diagram*{
						(v1) --[fermion]n1--[fermion](v2),
					};
				\end{feynman}
			\end{tikzpicture}
	}} + \ldots.
\end{align*}
We disregard the phonon dressing {of} the electromagnetic vertex
\[
\vcenter{\hbox{
		\begin{tikzpicture}
			\begin{feynman}
				\vertex [label=](v0) at (-1,0);
				\vertex [label=](v1) at (-0.5,0);
				\vertex [label=](v2) at (0.5,0);
				\vertex [label=](v3) at (1,0);
				\vertex [label=](v4) at (0,0);
				\vertex [label=] (n0) at (0,.8);
				\diagram*{
					(v0) --[fermion](v3),
					(v1) --[scalar, half right] (v2),
					(v4) --[photon] (n0),
				};
			\end{feynman}
		\end{tikzpicture}
}} 
\]
This contribution provides {a} renormalization of {the} effective electric charge in {the} material. 
The result of this renormalization is fixed on the Fermi surface 
\(e=e(\mu)\) (disregarding temperature \(T\ll|\mu|\)) as long as we calculate DC or low frequency conductivity
(in comparison to \(\mu\hbar^{-1}\), where \(\mu\sim 100\) meV as in \(ZrTe_5\)).

In a material, magnetic field \(\vec H\) differs  from  magnetic field  \(\vec B\) (in Russian literature the latter is called magnetic induction, and it represents the true microscopic magnetic field), due to the material back-reaction. However, the question of the relevant approximation is not affected by this difference{, unless we} deal with a ferromagnetic material.

\begin{figure}[ht]
	\begin{tikzpicture}
		\begin{axis}[
			axis lines = middle,
			ylabel = {\large \(\omega\)},
			xlabel = {\large \(p_z\)},
			xmin=-2.6, xmax=2.1,
			ymin=-1, ymax=1.1,
			ticks=none,
			samples=50,
			width=10cm, height=10cm,
			grid=none
			]
			
			\addplot [
			thick,
			only marks,
			mark=o, 
			mark size=1.5pt,
			mark options={solid, red},
			] table { 
				-1.9 0.64 
			};
			
			\addplot [
			samples=48,
			domain=-1.87:0.4,
			thick,
			only marks,
			mark=*, 
			mark size=1.5pt,
			mark options={solid, red},
			] {sin(deg(2*x))};
			\node[red] at (axis cs:0.4,0.4) {\Large R};
			\node[red] at (axis cs:-1.6,0.4) {\Large \(\bar L\)};
			
			\addplot [
			dashed,
			domain=-3:.4,
			] {0.68};
			\draw[<->, thick] (axis cs:-2.5,0) -- (axis cs:-2.5,0.68) node[midway, right] {\Large \(\mu\)};
            \draw[->, thick] (axis cs:-0.3,-0.9) -- (axis cs:0.3,-0.9) node[midway, above] {\Large \(\vec{E},\vec{B}\)};
			\node[black] at (axis cs:0.2,-0.1) {\Large \(\Gamma\)};
			\node[black] at (axis cs:-1.7,-0.1) {\Large \(\bar\Gamma\)};
			\draw[dashed] (axis cs:3.14/4,-3) -- (axis cs:3.14/4,3);
			\node[black] at (axis cs:3.14/2+0.2,-0.1) {\Large \(\bar\Gamma\)};
			
			\draw[->, thick, red] (axis cs:0.40,0.73) arc[start angle=40, end angle=250, radius=5pt];
			\draw[->, thick, red] (axis cs:-1.85,0.57) arc[start angle=300, end angle=100, radius=5pt];
			\draw[decorate, decoration={snake, amplitude=1mm, segment length=5mm}, thick, ->]
			(axis cs:0.26,0.73) -- (axis cs:-0.51,0.9) node[draw, ellipse, left, align=center] {Phonons with \\ temperature \(T\)};
			
			\addplot [
			thick,
			only marks,
			mark=o, 
			mark size=1.5pt,
			mark options={solid, blue},
			] table { 
				-0.33 0.64 
			};
			\addplot [
			samples=50,
			domain=-0.3:1.97,
			thick,
			only marks,
			mark=*, 
			mark size=1.5pt,
			mark options={solid, blue},
			] {sin(deg(-2*x))};
			\node[blue] at (axis cs:-0.4,0.4) {\Large L};
			\node[blue] at (axis cs:1.5,0.4) {\Large \(\bar R\)};
			\draw[<->, thick] (axis cs:-2.5,0) -- (axis cs:-2.5,-1) node[midway, right] {\Large \(\Lambda\)};
			\addplot [
			dashed,
			domain=-.4:.8,
			] {0.63};
			\addplot [
			dashed,
			domain=-.4:.8,
			] {0.73};
			\draw[<->, thick] (axis cs:0.8,0.63) -- (axis cs:0.8,0.73) node[midway, right] {\(\sim\vec{E}\vec{B}\ep_B^{-1}\)};
			\draw[->, thick, blue] (axis cs:1.97,0.73) arc[start angle=40, end angle=250, radius=5pt];
			\draw[->, thick, blue] (axis cs:-0.29,0.57) arc[start angle=300, end angle=100, radius=5pt];
			\draw[decorate, decoration={snake, amplitude=1mm, segment length=5mm}, thick, ->]
			(axis cs:1.8,0.73) -- (axis cs:1.1,0.9) node[midway, above right] {phonon};
			
		\end{axis}
	\end{tikzpicture}
	\caption{{
		A sketch of the Dirac band population density in the external electric and magnetic fields for Dirac semimetals in the limiting case of vanishing mass parameter $m$.
		The external electric field \(\vec{E}\) pumps up the chiral imbalance via the chiral anomaly. The interplay of chiral symmetry violation (due to the lattice regularization with physical cut-off \(\Lambda\))
		and the energy relaxation due to the interaction with the thermal bath of phonons (resulted in the  dissipation rate \(\ep_B\sim T\)) and with impurities
         provides the relaxation mechanism.  
         This  process is inverse to the pumping of electron-hole pairs by the chiral anomaly.
		The \(\bar\Gamma\) points are identical and the regions on the plot are to be overlayed.
		\label{FigBrZ}}
	}
\end{figure}

\subsection{Linear response to uniform electric field}
\label{SectIIIB}

In \cite{Banerjee2020obs} the general expression has been derived for the  electric current density  in Keldysh technique {(see also the earlier development of the same formalism in  \cite{shitade2017anomalous,onoda2006theory,onoda2006intrinsic,sugimoto2007gauge,onoda2008quantum})}: 
\begin{eqnarray}
J^k(x) & = & {-}
\frac{i e }{2}\int \frac{d^{{D+1}}\pi}{(2\pi\hbar)^{{D+1}}} \tr\left((\partial_{\pi_{k}}\hat{Q}_W)\hat{\bf G}_W\right)^{<}
\\&&
{-}\frac{i e}{2}\int \frac{d^{{D+1}}\pi}{(2\pi \hbar)^{{D+1}}}
\tr\left(\hat{\bf G}_W (\partial_{\pi_{k}}\hat{Q}_W)\right)^{<}\nonumber\\
& = & {-}
\frac{i e }{2}\int \frac{d^{{D+1}}\pi}{(2\pi\hbar)^{{D+1}}} \tr\left((\partial_{\pi_{k}}\hat{Q}^R_W)\hat{\bf G}^<_W\right)
\\&&
{-}\frac{i e}{2}\int \frac{d^{{D+1}}\pi}{(2\pi \hbar)^{{D+1}}}
\tr\left(\hat{\bf G}^<_W (\partial_{\pi_{k}}\hat{Q}^A_W)\right), \, k = 1,2,3.\nonumber
\end{eqnarray}
Here $\pi = p - e A(x)$,  $D$ is dimension of space (in our case $D=3$), 
$G_W$ and $Q_W$ are the Wigner transformed Keldysh Green function and its inverse (without interactions). 
${\bf G}$ is the electron Green function with interactions taken into account through the self-energy operator (see Section \ref{SectSelf}). 

The electric charge density is expressed as 
\begin{eqnarray}
	J^0(x) & = & {-}
	\frac{i e }{2}\int \frac{d^{{D+1}}\pi}{(2\pi\hbar)^{{D+1}}} \tr\left((\partial_{\pi_{0}}\hat{Q}^R_W)\hat{\bf G}^<_W\right)\nonumber\\&&
	{-}\frac{i e}{2}\int \frac{d^{{D+1}}\pi}{(2\pi \hbar)^{{D+1}}}
	\tr\left(\hat{\bf G}^<_W (\partial_{\pi_{0}}\hat{Q}^A_W)\right),    
	\label{J Wigner}
\end{eqnarray}
where, in contrast to the current density formula, 
    Green function and derivative of Dirac operator 
    cannot be placed under the common symbol $<$.

Similarly, the axial current and charge densities read (\(\mu=0,1,2,3\))
\begin{eqnarray}
	J_5^\mu(x) & = & {-}
	\frac{i }{2}\int \frac{d^{{D+1}}\pi}{(2\pi\hbar)^{{D+1}}} \tr\left(\gamma^5(\partial_{\pi_{\mu}}\hat{Q}^R_W)\hat{\bf G}^<_W\right)\nonumber\\&&
	{-}\frac{i }{2}\int \frac{d^{{D+1}}\pi}{(2\pi \hbar)^{{D+1}}}
	\tr\left(\gamma^5\hat{\bf G}^<_W (\partial_{\pi_{\mu}}\hat{Q}^A_W)\right).
	\label{J5 Wigner}
\end{eqnarray}
{\it 
    In the above expressions the non-renormalized axial and vector currents are considered. 
    In the presence of interactions the Green function receives self-energy correction $\Sigma$ (see Section \ref{SectSelf}). 
    Correspondingly, one can define its inverse $\hat {\bf Q}$, which receives corrections as well ${\hat {\bf Q}} = {\hat Q} - {\hat \Sigma}$. 
    These corrections, however, do not enter the above expressions for $J$ and $J_5$.
}

The linear response of Keldysh--Green function to external electromagnetic field strength is given by \cite{Banerjee2020obs,shitade2017anomalous,onoda2006theory,onoda2006intrinsic,sugimoto2007gauge,onoda2008quantum}
\begin{equation}
	\hat{\bf G}^{(1)}_W ={-}  \frac{ i eF^{\mu\nu}}{2}\hat{\bf G}_W \star \partial_{\pi^{\mu}}\hat{\bf Q}_W  \star\hat{\bf G}_W \star \partial_{\pi^{\nu}}\hat{\bf Q}_W \star \hat{\bf G}_W ,
	\label{QGK1}
\end{equation}
where the star product is defined in Eq.\eqref{EqStarProd}.
Substituting Eq. (\ref{QGK1}) to the expressions for vector and axial currents,
    we will obtain their response to electric field.

\subsection{Relation between chiral density and electric conductivity}

{Substituting Eq. (\ref{QGK1}) to Eq. (\ref{J5 Wigner}) we obtain expression for the chiral density
	\begin{eqnarray}
		\rho_5(x) & = & {-} \frac{  e E_j}{2}
		\Re  \int \frac{d^{{4}}\pi}{(2\pi\hbar)^{{4}}} \tr\Bigl(\gamma^5   \hat{\bf G}_W \star \partial_{\pi_{[0}}\hat{\bf Q}_W \nonumber\\ && \star\hat{\bf G}_W \star \partial_{\pi_{j]}}\hat{\bf Q}_W \star \hat{\bf G}_W  \Bigr)^{<}(\partial_{\pi_{0}}{Q}^A_W)
		\end{eqnarray}
	Here only the last multiplier $\hat{Q}$ does not contain interaction corrections, while the preceding $\bf G$ and $\bf Q$ represent the complete interacting Green function and its inverse. 
\begin{eqnarray}
	\rho_5(x) & = & {-} \frac{  e E_j}{2}
	\Re  \int \frac{d^{{4}}\pi}{(2\pi\hbar)^{{4}}} \tr\Bigl(\gamma^5   \hat{\bf G}_W \star \partial_{\pi_{[0}}\hat{\bf Q}_W \nonumber\\ && \star\hat{\bf G}_W \star \partial_{\pi_{j]}}\hat{\bf Q}_W \star \hat{\bf G}_W  \Bigr)^{<}(\partial_{\pi_{0}}{ Q}^A_W)\nonumber\\ & = & 
	{-} { 4 e E_3 }
	\Re  \int \frac{dp_0 dp_3}{(2\pi\hbar)^{2}} \tr\Bigl(\gamma^5  {\tilde {\bf G}}  O^- \partial_{p_{[0}}\hat{\bf Q}_W \nonumber\\ && {\tilde { {\bf G}}}  O^- \partial_{p_{3]}}\hat{\bf Q}_W  {\tilde { {\bf G}}}  O^-\Bigr)^{<} (\partial_{p_{0}}\hat{ Q}^A_W)\nonumber\\&&
	\int \frac{dp_1 dp_2}{(2\pi\hbar)^{2}}  e^{-\frac{\pi_\bot^2}{|{e\hbar}B|} }\star  e^{-\frac{\pi_\bot^2}{|{e\hbar}B|} }\star  e^{-\frac{\pi_\bot^2}{|{e\hbar}B|} }	
\end{eqnarray}
One can check that in the chosen gauge the star product of the two exponentials in the above expression is equal to the same exponential:
\begin{equation}
2e^{-\frac{\pi_\bot^2}{|{e\hbar}B|} }\star  2e^{-\frac{\pi_\bot^2}{|{e\hbar}B|} } = 2e^{-\frac{\pi_\bot^2}{|{e\hbar}B|} }\label{ee}	,
\end{equation}
which relies on the fact that the stationary phase approximation provides the exact solution 
(see Appendix \ref{AppB}).
Then
\begin{eqnarray}
	\rho_5(x)  & = & 
	{-} \frac{  e E_3 |eB| }{4 \pi \hbar}
	\Re  \int \frac{dp_0 dp_3}{(2\pi\hbar)^{2}} \tr\Bigl(\gamma^5 O^-  {\tilde { {\bf G}}} \partial_{p_{[0}}{\tilde { {\bf Q}}} \nonumber\\ && {\tilde {\hat {\bf G}}}  \partial_{p_{3]}}{\tilde { {\bf Q}}}  {\tilde {{\bf G}}} \Bigr)^{<} (\partial_{p_{0}}{\tilde Q}^A),
	\label{rho5 Wigner}
\end{eqnarray}
where ${\tilde { {\bf Q}}} = {\tilde { {\bf G}}}^{-1}$. 

For the electric current we obtain in a similar way
\begin{eqnarray}
	J^k(x)  & = & 
	{-} \frac{  e^2 E_3 |eB| \delta^{k3}}{ 4\pi \hbar}
	\Re  \int \frac{dp_0 dp_3}{(2\pi\hbar)^{2}} \tr\Bigl( O^-  {\tilde { {\bf G}}} \partial_{p_{[0}}{\tilde { {\bf Q}}} \nonumber\\ && {\tilde { {\bf G}}} \partial_{p_{3]}}{\tilde { {\bf Q}}} {\tilde { {\bf G}}} \Bigr)^{<} (\partial_{p_{3}}{\tilde Q}^A),
	\label{j Wigner}
\end{eqnarray}

The following property of the gamma matrices
\begin{equation}
	\gamma^5 O^- = - {\rm sign}\,(e B) \alpha^3 O^-,
\end{equation}
and taking into account that $\partial_{p_{3}}{\tilde Q}^A = v_F \alpha_3$, $\partial_{p_{0}}{\tilde Q}^A=1$,
we obtain
\begin{equation}
	J^k = - e v_F {\rm sign} (e B) \delta^{k3} \rho_5.\label{Jrho}
\end{equation}
}
{\it Thus the electric current is related rigidly to the chiral imbalance irrespective of the nature of the chirality relaxation mechanism and the way of energy dissipation. This is the same relation as the one proposed by the CME conjecture (compare Eq. (\ref{EqRho5CME}) and Eq. (\ref{EqCond2CME})).} 

\subsection{Calculation of chiral density}

{Now we are in the position to calculate the chiral density
\begin{eqnarray}
	\rho_5  & = & 
 \frac{  e^2 (\vec{E} \vec{B}) }{4 \pi v_F \hbar}
	\Re  \int \frac{dp_0 dp_3}{(2\pi\hbar)^{2}} \tr\Bigl( O^-  {\tilde { {\bf G}}}  \partial_{p_{[0}}{\tilde { {\bf Q}}} \nonumber\\ && {\tilde { {\bf G}}} \partial_{p_{3]}}{\tilde { {\bf Q}}} {\tilde G}  (\partial_{p_{3}} {\tilde Q})\Bigr)^{<}.
	\label{r5}
\end{eqnarray}
Using the theorem proven in Appendix A of \cite{Banerjee2020obs} we represent this expression as the sum of the contributions of the Dirac sea and of the Fermi surface $\rho_5 = \rho_5^{(sea)} + \rho_5^{(FS)}$:
\begin{eqnarray}
	\rho^{(sea)}_5  & = & 
	\frac{  e^2 (\vec{E} \vec{B}) }{ 4\pi v_F \hbar}
    \Re  \int \frac{dp_0 dp_3}{(2\pi\hbar)^{2}} \tr\Bigl( O^-  {\tilde { {\bf G}}}^A  \partial_{p_{[0}}{\tilde { {\bf Q}}}^A \nonumber\\ 
        && {\tilde { {\bf G}}}^A  \partial_{p_{3]}}{\tilde { {\bf Q}}}^A  {\tilde { {\bf G}}}^A \partial_{p_{3}}{\tilde {\bf Q}}^A\Bigr) {n(p_0)} 
	\label{r5sea}
\end{eqnarray}
and
\begin{eqnarray}
	\rho^{(FS)}_5  & = & \frac{  e^2 (\vec{E} \vec{B}) }{ 4\pi v_F \hbar}
        \Re  \int \frac{dp_0 dp_3}{(2\pi\hbar)^{2}} n'(p_0) \tr\Bigl(O^-  \nonumber\\ 
    &&\quad{\tilde { {\bf G}}}^R  ({\tilde { {\bf {Q}}}}^A - {\tilde { {\bf Q}}}^R)  
        {\tilde { {\bf G}}}^A  \partial_{p_{3}}{\tilde { {\bf Q}}}^A  {\tilde { {\bf G}}}^A  
        \partial_{p_{3}}{\tilde Q}^A\Bigr)\nonumber\\
	&&-\frac{  e^2 (\vec{E} \vec{B}) }{ 4\pi v_F \hbar}
	    \Re  \int \frac{dp_0 dp_3}{(2\pi\hbar)^{2}} n'(p_0) \tr\Bigl( O^- \nonumber\\  
    &&\quad {\tilde { {\bf G}}}^R  \partial_{p_{3}}{\tilde { {\bf Q}}}^R  
        {\tilde { {\bf G}}}^R  ({\tilde { {\bf Q}}}^A - {\tilde { {\bf Q}}}^R)  
        {\tilde { {\bf G}}}^A  \partial_{p_{3}}{\tilde Q}^A\Bigr)\nonumber\\
	& = & \frac{  e^2 (\vec{E} \vec{B}) }{ 4\pi v_F \hbar}
	    \Re  \int \frac{dp_0 dp_3}{(2\pi\hbar)^{2}} n'(p_0) \tr\Bigl( O^-  \nonumber\\ 
    &&\quad \partial_{p_{3}}({\tilde { {\bf G}}}^A  +{\tilde { {\bf G}}}^R) 
        \partial_{p_{3}}{\tilde { { Q}}}^A\Bigr)\nonumber\\
	&&+ \frac{  e^2 (\vec{E} \vec{B}) }{ 4\pi v_F \hbar}
	    \Re  \int \frac{dp_0 dp_3}{(2\pi\hbar)^{2}} n'(p_0) \tr\Bigl( O^-  \nonumber\\ 
    &&\quad {\tilde { {\bf G}}}^R   \partial_{p_{3}} 
        ({\tilde { {\bf Q}}}^A + {\tilde { {\bf Q}}}^R)  
        {\tilde { {\bf G}}}^A  \partial_{p_{3}}{\tilde Q}^A\Bigr).
	\label{r5FS}
\end{eqnarray}
The Dirac sea term vanishes identically due to the cyclic property of trace, 
    provided that the integrals are convergent. 
The convergence in infrared is provided by the energy dissipation rate $\epsilon_B$ and by the small fermion mass $m$. 
The given integral is convergent in ultraviolet due to the behavior of Dirac operator at large momenta. 
However, ultraviolet regularization of the theory is necessary in order to provide finite values for the other physical quantities,
    as well as to explain the chirality relaxation process. 
The most natural choice is the lattice regularization, 
    which reflects the nature of electrons in solids. 
We do not specify here the regularization because the result does not depend on its choice.  
{\it 
    We also take into account that 
    $\partial_{p_3} {\tilde { {\bf Q}}}^{A/R} = \partial_{p_3}{\tilde { { Q}}}^{A/R}$, 
    which is justified by our calculation of electron self-energy $\Sigma$ 
    (see Section \ref{SectSelf}).
} 
Namely, the explicit expression for $\hat{\Sigma}$ with contributions of disorder and interaction with phonons is 
\begin{eqnarray}
	\hat{\Sigma} &=& \frac{1 + \frac{m v_F^2}{p_0}\gamma^0}{\sqrt{1-\frac{{m^2 v_f^4}}{p_0^2}}} \left(\begin{array}{cc}
	- i \epsilon_B & 2 i \epsilon_B n(p_0) \\0 &  i \epsilon_B 	
    \end{array} \right), {|p_0|}>m v_F^2\nonumber\\ 
\hat{\Sigma} &=& \frac{1 + \frac{m v_F^2}{p_0}\gamma^0}{\sqrt{-1+\frac{{m^2 v_f^4}}{p_0^2}}} \left(\begin{array}{cc}
	-  \epsilon_B & 2 i \epsilon  n(p_0) \\0 &  - \epsilon_B 	
\end{array} \right), {|p_0|}\le m v_F^2
\end{eqnarray}
with $\epsilon \to 0$ and
\begin{equation}
	\epsilon_B = (\lambda + \chi)\,\mu.
\end{equation}
$\lambda$ and $\chi$ are numerically small factors (see Eqs.\eqref{EqEpsChi},\eqref{EqEpsLmbd},\eqref{EqChiLambdaZrTe5}), 
    which  depend on $B$, $T$, and constants.

With the notations similar to those of Eq. (\ref{wH1})
\begin{gather}
	{\bf w}_\pm=p_0\pm i\hat{\ep}_B, \quad {\bf m}_\pm=m (1 \pm i\hat{\ep}_B/p_0),  \nonumber\\{\bf w}=p_0, \quad \tilde {\bf G}^{R(A)} = \hbar({\bf w}_\pm -\tcH_\mp)^{-1}, \nonumber\\
	\tcH_\pm = v_F\al_3p_3 + \gamma_0 {\bf m}_\pm v_F^2, \quad
	\tE_\pm=\sqrt{v_F^2p_3^2+{\bf m}_\pm^2}, \\
	 \tilde{\bf G}^< = (\tilde{\bf G}^A - \tilde{\bf G}^R)n({\bf w}),\,
	 {\hat{\ep}_B = \frac{\ep_B}{\sqrt{1-\frac{m v_f^2}{\mu^2}}}}, \nonumber
\end{gather}
in the relevant limit \(|\mu|\gg T,m,\hat{\ep}_B\) the Fermi distribution is nearly degenerate, 
    and the integral over \(p_0\) is trivial
\begin{align}
	{\rho^{(FS)}_5} &=  
	    \frac{ { v_F e^2} (\vec{E} \vec{B}) }{4 \pi  \hbar}
        \Re  \int \frac{ dp_3}{(2\pi\hbar)^{2}}  \tr\Bigl(  
        \tilde{\bf G}^R   \alpha^3 \tilde{\bf G}^A  \alpha^3 \Bigr){\Big|_{p_0=\mu}}\nonumber\\ 
    &=\frac{{e^2 \hbar v_F} (\vec{E} \vec{B})}{4 \pi }\Re\int {\frac{dp_3}{(2\pi\hbar)^2}}\nonumber\\
    &\quad\times\tr \al^3\frac{{\bf w}_++\tcH_-}{{\bf w}_+^2-\tE_-^2}
        {\al^3}\frac{{\bf w}_-+\tcH_+}{{\bf w}_-^2-\tE_+^2}{\Big|_{p_0=\mu}}\nonumber\\ 
    &= \frac{e^2   (\vec{E} \vec{B})}{{4}\pi^2 \hbar  \hat{\ep}_B }
        \frac{\sqrt{1- m^2 v_F^4/\mu^2}}{1+ m^2 v_F^4/\mu^2}.\label{tau5MR}
\end{align}
The chirality relaxation time appeared in our calculation naturally 
(c.f. the latter to Eq.\eqref{EqRho5CME})
\begin{eqnarray}
    \tau_5 &=& \frac{{\hbar}}{{2}\hat{\ep}_B}\frac{\sqrt{1- m^2 v_F^4/\mu^2}}{1+ m^2 v_F^4/\mu^2}=\frac{{\hbar}}{{2}{\ep}_B(1+ m^2 v_F^4/\mu^2)}. \label{EqTau5}
\end{eqnarray}
Correspondingly, the conductivity receives the form identical to Eq.\eqref{EqCond2CME}
\begin{align}
    \sigma_{kj} \approx \frac{|e|^3 v_FB_jB_k}{{2\pi^2}\hbar^2 |{\vec B}|}\tau_5. \label{EqSigmaH}
\end{align}

\subsection{Discussion of the zero gap limit}\label{SectCondZG}

The last expression is well defined even in the limit $m=0$. 
A conventional Drude contribution arising from microscopic particle collisions can remain finite even with zero Dirac mass. 
Therefore, it follows from our Eq. (\ref{Jrho}) that the chiral density remains finite as well in the massless case (since it will be given by the Drude formula). 
The reason for this is the breakdown of chiral symmetry in ultraviolet. 
The situation here is {somehow} similar to that of the chiral anomaly in quantum field theory. 
In the theory that is precisely chiral symmetric (on the level of lagrangian) the ultraviolet regularization breaks the chiral symmetry, and results in the nonzero value of the divergence of axial current. 
In particular, within the lattice regularization (most close to the situation in solids) the explicit breakdown of chiral symmetry occurs far from the Fermi points, i.e. in the region of momentum space situated far in the ultraviolet region. 
\hl{Same results were obtained in \cite{Lu_2015,Li_2023}, and application of the Drude formula was discussed there.}

In our paper we consider  the  effective low energy theory for the electronic quasiparticles in Dirac semimetals. This is the continuous theory of Dirac fermions. The more fundamental theory of these quasiparticles has the form of a certain lattice regularization of the mentioned continuous theory. Therefore, in the mentioned microscopic theory (i.e. in the specific lattice regularization of the considered continuous theory) the chiral symmetry is broken in the ultraviolet, even if the mass parameter of Dirac fermions is zero.

{Let us discuss here specifically the case $m=0$.  
The obtained expression for the chiral density of Eq. (\ref{rho5 Wigner}) is equal to the {\it difference} between the two terms: the density of the left-handed electrons and the density of the right-handed fermions. (This difference is encoded in the $\gamma^5$ matrix standing inside the trace. The trace itself is therefore the difference of the left-handed and the right-handed terms.) If we suppose for a moment that, say, the right-handed electrons exist alone, then the existence of a finite dissipation rate $\epsilon_B$ would result in the excess of the number of particles over the equilibrium value given by the half of the value of Eq. (\ref{tau5MR})\footnote{Notice, that the equilibrium number of electrons itself (that is the number of electrons within the Dirac sea) is divergent and needs regularization.}.  This would mean that the total number of electrons is not conserved. Of course this is not correct because the right-handed electrons cannot exist in solids without the left-handed ones, which corresponds to the Neielsen--Ninomiya theorem. The excess of the left-handed fermions over their equilibrium number is equal to the excess of the right handed ones, but with the opposite sign. This provides the conservation of electric charge, and results in the total chiral density given above in Eq. (\ref{tau5MR}). This pattern is 
illustrated by Fig. \ref{FigBrZ}. Let us have in mind a lattice regularization, in which the effective one-dimensional Hamiltonian of Eq. (\ref{wH1}) (for $m=0$) and the corresponding energy levels are regularized as
\begin{gather}
	\tcH = v_F\al_3 \frac{{\rm sin}\,p_3 a}{a} + v_F \gamma_0 \frac{(1-{\rm cos}\,p_3 a)}{a}, \nonumber\\ 
	\tE=\pm \frac{v_F}{a} {\rm sin}\,\frac{p_3 a}{2}. \label{wH2}
\end{gather}
Here parameter $a$ means the lattice spacing, which tends to zero in continuous limit.
Then the left-handed branch of spectrum is connected to the right-handed branch through the boundary of the Brillouin zone. Therefore, the lack of particles on one branch results in their excess on the other branch. Chiral anomaly produces motion of occupied states, due to which the excess of the right-handed electrons and the lack of the left-handed ones (or, vice-versa) grows linearly in time. In the first order of Keldysh perturbation theory considered here we then arrive at the divergent expression for the chiral density (the one of Eq. (\ref{tau5MR}) with $\epsilon_B \to 0$). The presence of finite value of $\epsilon_B$ (as it was discussed above at the end of Sect. \ref{SectIIIB}) means that inside each branch (left handed or right handed) the "occupied" states are actually not stable, that is the corresponding particles may jump into the other states. In particular, this means that the jumps occur with directions pointed out on Fig. \ref{FigBrZ} .  This is the counter-flow of states that compensates partially the flow resulted from the chiral anomaly. As a result, the finite value of the chiral density given by Eq. (\ref{tau5MR}) appears.  One can see that this is the presence of the finite dissipation rate, which provides the relaxation of chirality. At the same time, transition of the left-handed electrons into the right-handed ones (and vice versa) occurs even at $m=0$ due to the regularization. In lattice regularization the left-handed branch of spectrum is continued across the boundary of Brillouin zone as the right-handed branch. 
However, the same conclusion can be drawn on the basis of the another regularization. One can use also, for example, the Pauli--Villars one. In that case the mass parameter of the Pauli--Villars regulator breaks chiral symmetry explicitly, which gives the same result.   
}

{In this respect we would like to mention the other phenomenon (not related directly to axial anomaly, but related to the {way how  regularization may modify the values of physical quantities}).  Let us discuss 2+1 D relativistic Dirac fermions in the presence of magnetic field. Such a system is realized in the real material – the graphene monolayer. The standard calculation of the QHE conductivity gives the standard result – the conductivity is given by inverse Klitzing constant times the number of occupied Landau levels. The naive non – regularized theory contains infinite number of occupied Landau levels situated below zero (below half filling in graphene). The existence of these levels is intimately related to the Dirac sea (of the same theory without magnetic field). The answer for the QHE conductivity would become infinite in this non – regularized theory. Of course, this is not the correct answer. The ultraviolet regularization changes the pattern drastically. The Pauli--Villars regularization subtracts immediately the contribution of the Dirac sea, and the correct value of the QHE conductivity is obtained – we should count the number of occupied Landau levels starting from the zero level (the half filling in case of graphene) \cite{gusynin2005unconventional}. Of course, this is the answer that is experimentally verified in real materials. Even more interesting situation occurs in lattice regularization (i.e. if we come back to the tight – binding model of graphene from its low energy continuous field theoretic description). Namely, the answer for the QHE conductivity is given by the sum of Chern numbers for the occupied Landau levels (the sum includes the levels situated below zero). It appears that some of the low lying Landau levels have large negative  Chern numbers. These cancel precisely the remaining positive Chern numbers of the Landau levels situated below zero. For the detailed description of this interesting phenomenon see \cite{hatsugai2006topological}
		and related publications. 
    We again come to the conclusion that in the effective low energy theory the number of occupied Landau levels should be counted from zero. We consider that this phenomenon is in some sense a cousin of our vanishing contribution to the axial density originated from the Dirac sea.     
}

{It is worth mentioning that the chirality relaxation time calculated in \cite{Boyarsky_2021} is essentially different from the one calculated in our paper. Namely, in \cite{Boyarsky_2021} the charge $Q_5$ is considered, which includes both the ordinary axial charge and the contribution resulted from the Chern--Simons term. Together, these two contributions form the conserved quantity in the limit of massless theory (even with the ultraviolet regularization taken into account). This is why the chirality relaxation rate calculated in \cite{Boyarsky_2021} vanishes in the massless limit.       }

\section{Electron self-energy}
\label{SectSelf}

In this section we calculate the level width that defines the relaxation time of Eq.\eqref{EqTau5epBc} 
    and hence magnetoconductivity and axial charge density.
The level half-width is parametrized with imaginary contribution to time components of the fermion propagator, 
    which is pure regularization \(\ep\to +0\) for free fermions.
The finite value of the imaginary part is provided by the fermion self-energy.
We calculate the contributions from interactions with acoustic longitudinal phonons and simple non-magnetic disorder.

\subsection{Contribution to the electron self-energy due to the interactions with acoustic phonons}\label{SectElPh}

In this subsection we suggest a model for the electron self-energy due to  interactions with acoustic phonons. 
We neglect effect of transverse acoustic phonons as well as optical phonons, and concentrate on the interaction with the longitudinal acoustic phonons.

The interaction Hamiltonian reads (see Part 2, \S64 of \cite{LL9})
\begin{align}
    \hat H_\text{e-ph} &= \int d^{3}x ~\frac{w\hat\rho'}{\rho_0} 
        ~\hat\psi^\dag\hat\psi \label{EqHeph},
\end{align}
where \(w\) is the energy of volume deformation of the crystal lattice
(is of dimension of energy), 
\(\rho_0\) the unperturbed lattice density. Here, following \cite{LL9} we assume that phonons effectively may be considered as if they appear in liquid rather than in the crystal lattice. Correspondingly, we consider only the acoustic phonons that transport longitudinal vibrations. The other types of phonons, supposedly, do not change qualitatively the obtained {expression}  for the dissipation rate $\epsilon_B$.
{The interaction preserves the chiral symmetry.}

In contrast to relativistic QFTs, where the speed of light is the same for fermions and gauge bosons, 
in  topological semimetals 
    the Fermi velocity \(v_F\) 
    is few orders of magnitude larger then the sound velocity \(u\). 
$u$ is of the order of km/s as in any solid or liquid, 
    while $v_F$ in topological semimetals as well as in graphene is of the order of \(10^{-3}c\) \cite{CMEZrTe5,Crassee2018}.
In particular, in \(Cd_3As_2\) \(v_F\sim 10^5~ m/s, 
~ u\sim 10^3~ m/s, 
~ u/v_F \sim 10^{-2}\) 
\cite{Crassee2018,Neupane_2014, Liang2022ultrafast, Wang2007computation}.

\(u/{v_F}\ll 1\) simplifies the calculations in a way that 
    the distribution-dependent terms in the Keldysh-Green functions 
    are time-independent 
    (see Problems to \S 92 in \cite{LL10})
\begin{align}
    D^R = D^{--}-D^{-+} = ~&\hbar\frac{\rho_0k^2}{(\w+i\epsilon)^2-u^2k^2},\label{EqDss}\\
    D^< = D^{-+} = -i\pi \hbar\frac{\rho_0 k}{u}(&N_\bk\delta(\w-uk) +\\
    &+ (1+N_{-\bk})\delta(\w+uk)) \nonumber\\
    \approx -i\pi \hbar\frac{\rho_0k}{u}&\coth\frac{\beta uk}{2}\delta(\w),
\end{align}
where \(k=\sqrt{\bk^2},~\w=k_0\),
and we assumed the initial thermal distribution of phonons 
\begin{gather}
    N_\bk = (e^{\beta uk}-1)^{-1} \label{eqPhDist}.
\end{gather}
Note that due to the relatively small sound velocity \(u{/v_F}\ll 1\), 
    the effective temperature for phonons \(\sim T{v_F}/u\) is much larger than for electrons.
Thus we consider the thermal distribution for phonons 
    and the degenerate distribution for electrons,
    as in the previous section.

{We denote $g = w/\rho_0$.} The one-loop correction reads
\(
    \hat G_1 = {\hbar^{-1}}\hat G\hat \Sigma\hat G,
\)
where \cite{Arseev2015}
\begin{align}
    \Sigma^<(x,y) &= i{\hbar^{-1}}g^2G^<(x,y)D^<(x,y),  \\
    \Sigma^R(x,y) &= i{\hbar^{-1}}g^2(G^R(x,y)D^<(x,y) +\\
        &\quad +G^<(x,y)D^R(x,y)+G^R(x,y)D^R(x,y)). \nonumber
\end{align}
The Dyson equation allows to sum up diagram subseries 
\begin{gather}
    { {\hat {\bf Q}}} = \hat Q - \hat\Sigma,
    \quad { {\hat {\bf G}}} = \hbar({ {\hat {\bf Q}}} - \hat \Sigma)^{-1}, \label{EqDyson}
\end{gather}

\label{SectEpsH} 

In the strong magnetic field limit Eq.\eqref{EqApprMuB} we consider the LLL propagator.
Besides, we use property of Eq. (\ref{ee}), 
    which allows us to express the series with the self-energy correction as
\begin{eqnarray}
    { {\hat {\bf G}}} &=& G_\bot  {\tilde{\bf G}} O^- = G_\bot  {\tilde G} O^- + {\tilde G}   \hat{\Sigma} G_\bot \hat{G} O^-/\hbar \nonumber\\ && + {\tilde G} \hat{\Sigma} {\tilde G}  \hat{\Sigma} G_\bot  {\tilde G} O^-/\hbar^2 + \ldots \label{prodQG}
\end{eqnarray} 
with 
\begin{equation}
    {\tilde{\bf G}}^{-1} = {\tilde G}^{-1} - \Sigma/\hbar,
\end{equation} 
where ${\tilde G}$ is the unperturbed reduced $2D$ Green function, 
    and $\hat{\Sigma}$ is the fermion self-energy. 

The full calculation of the one-loop phononic electron self-energy is given in Appendix \ref{AppA},
    the result is
\begin{eqnarray}
    {\Sigma}^{R}_\text{{ph}} &=&{{\Sigma}^{A\dag}_\text{{ph}}\approx}
    -\chi \mu  \Bigl(1  + \frac{mv_F^2}{p_0}\gamma^0\Bigr) \frac{{i}}{\sqrt{1-\frac{m^2v_F^4}{p_0^2}}}
    \\ 
	\Sigma^{<}_\text{{ph}} &=& \chi \mu  \Bigl(1  + \frac{mv_F^2}{p_0}\gamma^0\Bigr) \frac{2 i n(p_0)}{\sqrt{1-\frac{m^2v_F^4}{p_0^2}}}  \Theta(p^2_0-m^2 v_F^4) \nonumber
\end{eqnarray}
where 
\begin{equation}
	\chi = \frac{|e B| w^2 T}{ {4}\pi \hbar^2 u^2  v_F\rho_0 \mu} = \frac{w^2}{{4}\pi l_B^2 l_T u^2 \rho_0 \mu} ,
    \label{EqEpsChi}
\end{equation}
(see Eq.\eqref{EqChiLambdaZrTe5} for the numerical estimates in ZrTe$_5$),    
{in the limit
\begin{equation}
    \frac{u}{\fin{T}}\left(\frac{\hbar |e B|}{2\pi }\right)^{1/2}\ll 1, \label{EqApprBuT}
\end{equation}}
(see Eq.\eqref{EqKappaZrTe5} for the numerical estimates in ZrTe$_5$).    

Thus, the level half-width contribution from the phonons in the strong magnetic field can be estimated as
\(\ep^\text{ph}_{B}=\chi \mu \sim |eB|T\).

\subsection{Contribution to the self-energy due to scattering on impurities}
\label{SectImp}

In this section we calculate the electron self-energy due to scattering on impurities. 
The presence of impurities is modeled by potential $U(\vec{x}) = \sum_i u(\vec{x}-\vec{x}_i)$, 
    where {the} sum is over the positions $\vec{x}_i$ of the impurities. 
The simplest $\delta$-function potential $u(\vec{x}) = u_0 \delta(\vec{x})$ is considered. 
The perturbative corrections to the electron Green function then read
\begin{eqnarray}
    \hat{G}_{xy}^{\text{{imp}}} &\approx&  \hat{G}_{xy}  
        + \int dz \hat{G}_{xz}  \hat{U}(\vec{z}) \hat{G}_{zy}/ \hbar + ...  \label{GG} 
\end{eqnarray}
where in the triangle representation used in the present paper
\begin{equation}
	\hat{U} = 1 \otimes U(\vec{x}).
\end{equation} 

The second term here being averaged over the disorder (i.e. over the random positions of impurities) is estimated as 
\begin{eqnarray}
 \langle\int dz \hat{G}_{xz}  \hat{U}(\vec{z}) \hat{G}_{zy}/ \hbar \rangle &= &  u_0 \langle \sum_i \int dz \hat{G}_{x z_i}   \hat{G}_{z_i y}/ \hbar \rangle \nonumber\\ & \approx & u_0 n_\text{{imp}} \int dz \hat{G}_{x z}   \hat{G}_{z y}/ \hbar, 	
\end{eqnarray}
where $z_i = (z_0,\vec{z}_i)$, while $n_\text{{imp}}$ is density of impurities. One can see that this contribution results effectively in the extra term in the Hamiltonian
\(
	\tcH \to \tcH + u_0 n_\text{{imp}}
\)
which can be absorbed in redefinition of the chemical potential $\mu \to \mu + u_0 n_\text{{imp}}$. 

The next term of expansion in Eq. (\ref{GG}) after averaging over disorder gives
\begin{eqnarray}
	\hat{G}^{(2)}_\text{{imp}} &= &  u^2_0 \langle \sum_{ij}  \int dz dz' \hat{G}_{x z} \sum_{i}\delta(z-z_i) \hat{G}_{z z'}\\ && \sum_j \delta(z'-z_j) \hat{G}_{z' y}/ \hbar^2 \rangle \nonumber\\ & = & u^2_0  \int dz dz' \hat{G}_{x z} n^{(2)}_\text{{imp}}(z-z') \hat{G}_{z z'} \hat{G}_{z' y}/ \hbar^2.	\nonumber
\end{eqnarray}
Here the pair correlator of disorder is given by 
\begin{equation}
	n^{(2)}_\text{{imp}}(z-z') \equiv \langle \sum_i \delta(\vec{z}-\vec{z}_i)   \sum_j \delta(\vec{z}'-\vec{z}_j) \rangle
\end{equation} 
In this expression the averaging is over the different configurations of the coordinate sets $\{ ..., \vec{z}_1, \vec{z}_2, ... \}$. 
Assuming that the disorder correlation length is much smaller than $l_\mu$ we use an approximation
\begin{equation}
	n^{(2)}_\text{{imp}}(\Delta z) \approx n_\text{{imp}} \delta (\Delta \vec{z}),
\end{equation}
where the delta-function is three-dimensional, and the correlator is time-independent.
The corresponding contribution to the self-energy is
\begin{equation}
	\hat{\Sigma}_\text{{imp}}(x,y) = u^2_0 n_\text{{imp}} \int d^3\vec{x}\hat{G}_{x y} \delta^{(3)}(\vec{x} - \vec{y})/\hbar
\end{equation}
In the absence of electric field after Fourier transformation we obtain
\begin{gather}
	\Sigma^{R(A)}_\text{{imp}}(\w) = u^2_0 n_\text{{imp}}\frac{|e B|}{2\pi \hbar}\int \frac{dp_3}{2 \pi \hbar}(\w_\pm -\tcH)^{-1}, \nonumber\\
	\tcH = v_F\al_3p_3 + \gamma_0 m v_F^2, \quad
    \tE=\sqrt{v_F^2p_3^2+m^2{v_F^4}}, \nonumber\\ \Sigma^{<}_\text{{imp}}(\w) = 2\pi u^2_0 n_\text{{imp}} \frac{|e B|}{2\pi \hbar} \int \frac{dp_3}{2 \pi \hbar} i{(\w+\tcH)\Delta_{\ep_B}} n(\w). \nonumber
\end{gather}
The calculations of the these integrals are similar to the calculations of the integrals over momenta for the case of phonons presented in Appendix \ref{AppA}, but simpler. 
In case of $\Sigma^<$ we take the integral due to the delta function $\Delta_\epsilon$, while for the calculation of the integral in expression for $\Sigma^{R(A)}$ we use the residue theorem
\begin{equation}
	\frac{\Sigma_\text{{imp}}^{<}(p_0)}{2n(p_0)}=i \lambda(B,\mu)  \frac{1 + m v_F^2 \gamma^0/p_0}{\sqrt{1 - m^2 v_F^4/p_0^2}} \mu \Theta(p_0-m v_F^2),
\end{equation}
and 
$\Sigma_\text{{imp}}^{{R}} = \Sigma_\text{{imp}}^{A\dagger}$,
\begin{eqnarray}
    \Sigma_\text{{imp}}^{A}&=&	- i\lambda(B,\mu)  \frac{1 + \frac{m v_F^2}{p_0} \gamma^0}{\sqrt{1 - \frac{m^2 v_F^4}{p_0^2}}} \mu, {~|p_0|}>mv_F^2, \nonumber\\
    \Sigma_\text{{imp}}^{A}&=&	- \lambda(B,\mu)  \frac{1 + \frac{m v_F^2}{p_0} \gamma^0}{\sqrt{-1 + \frac{m^2 v_F^4}{p_0^2}}} \mu, {~|p_0|}\le mv_F^2
\end{eqnarray}
with {a} small parameter
\begin{equation}
	\lambda = \frac{|eB|}{4\pi \hbar^2 } \frac{u_0^2 n_\text{{imp}}}{\mu\fin{v_F}}
        = \frac{1}{4\pi l_B^2 \hbar } \frac{u_0^2 n_\text{{imp}}}{\mu\fin{v_F}}
        \label{EqEpsLmbd}
\end{equation}
The level half-width contribution from the disorder in the strong magnetic field can be estimated as
\(\ep^\text{imp}_{B}=\lambda \mu \sim |eB|n_\text{imp}\), \hl{the same as in \cite{Li_2023}.}

As for numerical estimates, for $u_{{0}}/a^3 \sim 40$ \fin{eV}, $n_\text{{imp}} \sim 10^{-3} a^{-3}$ with $a \sim 10^{-9}$ m we obtain for $\mu \sim 100 $ {meV} and $B \sim 10$ T: $\lambda \sim 7 \times 10^{-4}$. This parameter is to be compared with parameter $\chi$ responsible for the contributions due to interactions with phonons. It was estimated roughly for $ZrTe_5$ in Appendix \ref{AppA} at the same values of $\mu$ and $B$, and at $T = 50$ K  
\begin{equation}
    \chi \sim 2\times 10^{-3},\quad \lambda \sim 7 \times 10^{-4}, \quad\chi/\lambda\sim 3. \label{EqChiLambdaZrTe5}
\end{equation}
The chosen values of $w$, $u_0$, $n_\text{{imp}}$ {fit} the results summarized in Fig. \ref{FigRhoT} match experimental data.

\section{Conclusions and discussion}\label{SectCon}

We investigated relation between the magnetoresistance of Dirac semimetals and the chiral magnetic effect.
We avoided introducing the {concept} of chiral chemical potential  \(\mu_5\) (or, better to say, we justify its appearance in a rigorous way),
    and ad-hoc free parameters such as relaxation times. 
For example, the chiral chemical potential can yield potentially unstable classical equations of  electrodynamics \cite{Avdoshkin2014gpa}. 
The chiral kinetic theory, which became popular recent years, supports the heuristic calculation of relation between the CME and negative magnetoresistance. However, it usually lacks a specified chirality relaxation mechanism,
    and accounts for the chirality relaxation with a phenomenological relaxation time \cite{CMEZrTe5,Monteiro2015mea}.

We consider a rather simple model, in which electrons in {vicinities of  Fermi points in Dirac} semimetals are modeled by  Dirac fermions.
{We calculate the contributions to electron self-energy of interactions with longitudinal phonons and of scattering on impurities. 
The {imaginary} part of the self-energy is related to dissipation of energy and finite value of electric conductivity.}  
For the given model we calculate conductivity {and axial charge density} using the Keldysh technique \cite{KamenevBook,Arseev2015,Kamenev2005course}. 

For Dirac semimetal Cd$_3$As$_2$ at $T = 10 $K, with $\mu \sim 100$ meV and in relatively large $B = 10 $T we estimate $l_B < l_\mu \ll l_T \ll l_0$
    --- the limit of strong magnetic field is achieved at  $B=10$ T. 
However, specifically to this material our analysis cannot be applied because in this material there is a different known mechanism of magnetoresistance \cite{Liang_2014}. 
For the materials with smaller values of $\mu$ the strong magnetic field limit can be achieved even for relatively small values of $B$. 
In particular, for Weyl semimetal $Nb P$ \cite{niemann2017chiral}, 
    $\mu \sim 10 $meV and $v_F \sim {10^{-3}c}$, 
    and thus at  $ B \ge 1 $T already $ l_\mu \approx 10^{-7} > l_B \approx 6 \times 10^{-8}$ m 
    --- strong magnetic field limit is apparently achieved at $1$ Tesla.  
For \(Na_3Bi\)  \cite{xiong2015evidence}, $\mu \sim 30 $meV and $v_F \sim {10^{-3}}c$ (along one of directions), 
    and thus at $T=10$ K, $B = 5 $T we estimate $ l_\mu \approx 4\times 10^{-8} > l_B \approx 3 \times 10^{-8}$ m
    ---strong magnetic field limit is achieved at around $5$ Tesla. 
In Dirac semimetal \(Cd_3As_2\) with lowered Fermi energy \cite{li2016negative} the limit of strong magnetic field is achieved also for the values of magnetic field essentially lower than $10$ Tesla. 

For Dirac semimetal  $ZrTe_5$ \cite{CMEZrTe5}, with $\mu \sim 100 $meV, $v_F \sim c/300$, 
    and at  $ B \ge 5 $T we estimate $ l_\mu \approx 4 \times 10^{-8}{~m} > l_B \approx 3 \times 10^{-8}$ m, 
    --- strong magnetic field limit is apparently achieved at $B=5$ T.  

The inequalities $l_0 \gg l_{{\mu}}, l_T \gg l_B$ mean that scattering on impurities as well as interaction with phonons may be considered as perturbations. 
We take into account interaction with phonons {and scattering on impurities} that are responsible for the dissipation of energy, {electric conductivity} and for the finite value of $\rho_5$. 

{The CME hypothesis in the strong magnetic field limit $l_B < \sqrt{2}\,l_\mu$ predicts \cite{Kharzeev2017} 
    the axial charge density given by Eqs.\eqref{EqRho5CME} and \eqref{EqCond2CME} 
    with an unspecified phenomenological relaxation time $\tau_5$. 
We obtain exactly the same functional dependence of $\rho_5$ of Eq.\eqref{tau5MR} and \(\sigma_{jk}\) of Eq.\eqref{EqSigmaH} on magnetic and electric fields, 
    while the chirality relaxation time \(\tau_5\) emerges naturally 
    --- in complete  agreement with the CME-based approach. }

{However, with the Keldysh technique and our semi-realistic model for the chirality relaxation mechanism 
    (explicit chiral symmetry breaking with the Dirac mass {and UV regularization}, and energy dissipation {from interaction with thermal bath of phonons} {and scattering on impurities}),
    we are able to estimate the chirality relaxation time \(\tau_5\).
    The suggested relaxation time essentially depends on magnetic field in the strong field limit. 
    {The dependence of relaxation time on mass parameter disappears, which {suggests} that the leading order in this mechanism is played by the region of Brillouin zone situated far from the Fermi points, where chiral symmetry is broken.}}

For the conductivity and chiral imbalance we reproduce Eqs.\eqref{EqCond2CME} and \eqref{EqRho5CME}
(with the  approximations Eqs.\eqref{EqApprMum},\eqref{EqApprMuB},\eqref{EqApprBuT} relevant for \(ZrTe_5\),
but with the relaxation time depending on the energy dissipation rate $\epsilon_B$ 
\begin{equation}
	\tau_5 = \frac{{ \hbar}}{{2}\epsilon_B }.
 \label{EqTau5epBc}
\end{equation}}
that in our model has two contributions---from longitudinal acoustic phonons and from scattering on impurities
	\begin{eqnarray}
        &&\epsilon_B = \epsilon_B^\text{{imp}} + \epsilon_B^\text{{ph}},\\
		&&\epsilon_B^\text{{imp}} = \lambda\, \mu, \quad \lambda = \frac{u_0}{4\pi l_B^2 \hbar } \frac{u_0 n_{imp}}{\mu \fin{v_F}}\nonumber\\
		&&\epsilon_B^\text{{ph}} = \chi\, \mu, \quad \chi  = \frac{w^2}{{4}\pi l_B^2 l_T u^2 \rho_0 \mu}\nonumber,
	\end{eqnarray}
{see Eq.\eqref{EqChiLambdaZrTe5} for the numerical estimates in ZrTe$_5$.}    
\hl{We found, that the level width itself is proportional to the magnetic field strength.}
 
With our expression for the dissipation rate the resulting expression for the strong magnetic field conductivity is
\begin{gather}
    \sigma_{kj} \approx \frac{B_jB_k}{ |{\vec B}|^2}(\rho_\text{ph}'T+\rho_\text{{imp}})^{-1}, \nonumber\\
    \rho_\text{ph}'= \frac{{\pi}}{{N_f}\hbar e^2v_F^2}\frac{w^2}{\rho_0u^2}, \quad
		\rho_\text{{imp}} = \frac{u_0^2n_i {\pi}}{{N_f}\hbar e^2\fin{v^2_F}}. \label{sBB}
\end{gather}
The result suggests saturation of the anomalous conductivity in strong magnetic field 
    and inverse proportionality to temperature.
The chirality breaking mechanism is provided by 
    the region of Brillouin zone situated far from the Fermi points, 
    (where the chiral symmetry is broken naturally),
    and the conductivity and axial charge are finite even for vanishing Dirac mass.

    Meanwhile, the raw data for resistivity \(\rho_{zz}\) along magnetic field in \(ZrTe_5\), 
    presented in Fig.S1 of \cite{CMEZrTe5} (see the {\sl Methods --- Transport Measurements} section in the arXiv v1 version), 
    demonstrates similar dependence 
    in the temperature range 30-70K 
    \begin{figure}[ht]
        \center{\includegraphics[width=0.9\linewidth]{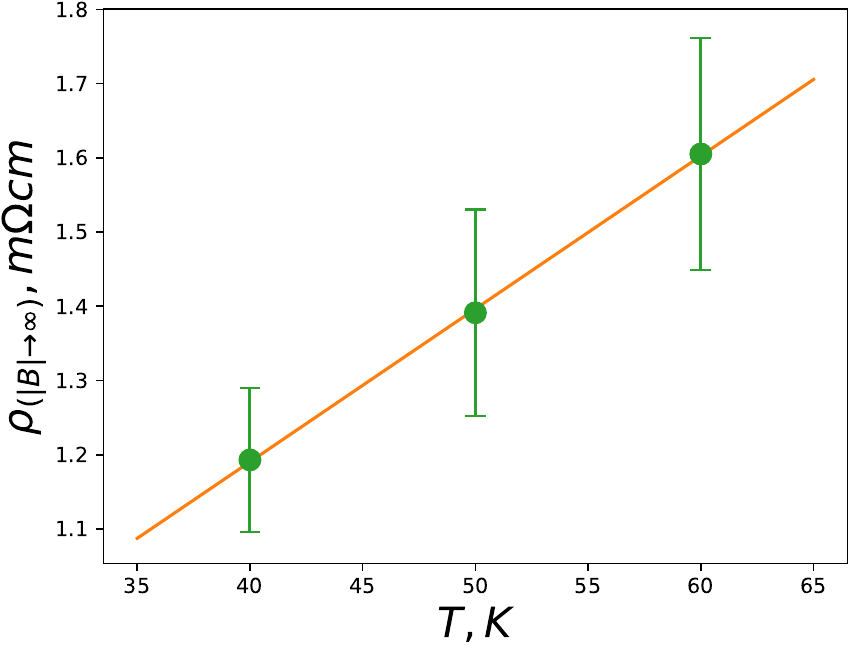}}
        \caption{Large magnetic field resistivity of \(ZrTe_5\).
            Data extracted (points) from Fig.S1 of \cite{CMEZrTe5} 
            (see the {\sl Methods --- Transport Measurements} section in the latest arXiv version)
            {as the average at \(B>6\) T and \(B<-6\) T,}
            fitted with the obtained here dependence Eq.\eqref{sBB}
            \(\rho_{zz} = \sigma_{zz}^{-1}=\rho_\text{{imp}}+\rho_\text{ph}'T\);
            fitting parameters values from the data are \(\rho_\text{{imp}}\sim0.37\pm0.10~m\Omega cm\), \(\rho_\text{ph}'\sim0.021~m\Omega cm K^{-1}\){, which roughly correspond to  values of our model parameters} $w\sim {1.3}$ eV, $u_0 \sim {30}$ \fin{eV} nm$^3$, $n_{imp} \sim   10^{-3}$ nm$^{-3}$. 
            \label{FigRhoT}
        }
    \end{figure}
    --- the resistivity saturates with the magnetic field and is approximately proportional to temperature. 
    At smaller temperatures, the non-monotonous magneto-conductivity is unexplained 
    -- apparently it is dictated by magnetic properties of the crystal lattice,
    {besides, our approximation Eq.\eqref{EqApprBuT} breaks down}. 
    At larger temperatures, the Lifshitz transition occurs, 
 which also might break our model assumptions.

\hl{It would be interesting to extend our analysis to the case of weak and intermediate magnetic fields (in experimental studies the magnetoresistance is typically investigated within the ranges of $0.1$ to $10$ T). 
A straightforward approach could be to calculate the magnetoconductivity with the complete sum over the Landau levels, as in \cite{ghosh2024anisotropic}, but for realistic semimetals models.
Alternative approach, which seems to be more feasible, 
    could be to derive a proper kinetic theory from the underlying QFT,
    as in \cite{Lin2019},
    and calculate observables with the kinetic theory.
It will then be interesting to check the validity of the CME conjecture in the general case.
}

\hl{One of the interesting questions is relation to topology of the coefficient standing in expression for the CME in Eq. (\ref{CME}). The similar expression for the chiral separation effect \cite{zubkov2023effect} contains a topological invariant composed of the two - point equilibrium Green  functions. A general expectation is that the CME conductivity is expressed through the same topological invariant. It is worth mentioning that such topological invariants reveal intimate relation between solid state systems and high energy physics  \cite{zubkov2018momentum,zubkov2012momentum,volovik2017standard} as well as with the fermionic superfluids \cite{volovik2013nambu}. It would be interesting also to extend our approach to the consideration of magnetoconductivity in quark matter, where the interactions are essentially non - perturbative, and various topological defects dominates dynamics (see, for example, \cite{bakker1999central} and references therein).}

\section*{Acknowledgements}

The authors are grateful to Igor Shovkovy for useful comments and pointing out us Eq. (24) of \cite{gorbar2014chiral}, which is reproduced in our work as well. The authors also kindly acknowledge useful comments by G.E. Volovik and Naoki Yamamoto.

\appendix
		
		\section{Proof of Eq. \ref{ee}}
		\label{AppB}
		In order to prove Eq. (\ref{ee}) one may use the basic property of Weyl symbols $A_W \star B_W = (\hat{A} \hat{B})_W$ (see \cite{Banerjee2020obs}), and consider transverse Green functions ${\hat G}_\bot$ given by 
		\begin{equation}
			\langle x | {\hat G}_\bot | z \rangle = 	\frac{|eB|}{2\pi \hbar}e^{i  \int_x^z eA(w)dw/\hbar - \frac{|e B|}{4\hbar}(\vec{x} - \vec{z})^2} 
		\end{equation}
		One can check easily that Weyl symbol of this operator is $G_{\bot,W} = 2 e^{-\frac{\pi_\bot^2}{|{e\hbar}B|} }$ with $\pi = p - e A(x)$, and for the matrix elements $\langle x |\hat{G}_\bot^2| y\rangle$ we obtain (assume here for simplicity $e>0$ and $B = |{\bf B}|$):
		\begin{eqnarray}
			&&\int d^2 z \frac{|eB|^2}{(2\pi \hbar)^2}{\rm exp}\,\Bigl(\frac{i e }{4\hbar}  (\vec{z} - \vec{x}) [(\vec{x}+\vec{z})\times \vec{B}] - \frac{e B}{4\hbar}(\vec{x} - \vec{z})^2\nonumber\\&&+\frac{i e }{4\hbar}  (\vec{y} - \vec{z}) [(\vec{y}+\vec{z})\times \vec{B}] - \frac{e B}{4\hbar}(\vec{y} - \vec{z})^2\Bigr)\nonumber\\ 
			&=& \frac{|eB|}{(2\pi \hbar)}{\rm exp}\,\Bigl(\frac{i e }{4\hbar}  (\vec{z} - \vec{x}) [(\vec{x}+\vec{z})\times \vec{B}] - \frac{e B}{4\hbar}(\vec{x} - \vec{z})^2\nonumber\\&&+\frac{i e }{4\hbar}  (\vec{y} - \vec{z}) [(\vec{y}+\vec{z})\times \vec{B}] - \frac{e B}{4\hbar}(\vec{y} - \vec{z})^2\Bigr)\Big|_{\vec{z} = \vec{z}_{ext}(\vec{x},\vec{y})}
			\nonumber\\ 
			&=& \frac{|eB|}{(2\pi \hbar)}{\rm exp}\,\Bigl(\frac{i e }{2\hbar}  \vec{B} [\vec{z} \times \vec{x}]  - \frac{e B}{4\hbar}(\vec{x}^2 + \vec{z}^2  - 2 \vec{z} \vec{x})\nonumber\\&&+\frac{i e }{2\hbar}  \vec{B}[\vec{y} \times \vec{z}] - \frac{e B}{4\hbar}(\vec{y}^2 + \vec{z}^2 - 2 \vec{z}\vec{y}\Bigr)\Big|_{\vec{z} = \vec{z}_{ext}(\vec{x},\vec{y})}
		\end{eqnarray}
		Here $\vec{z}_{ext}$ is the solution of equation that expresses the stationary phase condition:
		\begin{equation}
			\vec{z}_{ext} = \frac{\vec{x}+\vec{y}}{2} + i\frac{[(\vec{x} - \vec{y})\times \vec{B}]}{2B}.
		\end{equation}
		We arrive at
		\begin{eqnarray}
			&&	\langle x | \hat{G}_\bot^2 |y\rangle = \frac{|eB|}{(2\pi \hbar)}{\rm exp}\,\Bigl(\frac{i e }{2\hbar}  \vec{B} [\vec{y} \times \vec{x}]  - \frac{e B}{4\hbar}(\vec{x}- \vec{y})^2 \Bigr)
		\end{eqnarray}
		that is
		\begin{equation}
			\hat{G}_\bot^2 = \hat{G}_\bot \label{eeA}
		\end{equation}
	Eq. (\ref{ee}) appears as the Wigner transformation of this expression.
		
		\section{Contribution of interactions with phonons to electron self-energy }
		\label{AppA}

		\subsection{Calculation of the lesser component of self - energy}
		
		We assume that $B_3 >0$ and obtain for the Wigner transformation of its `lesser' component:
		{\begin{align}
				& \Sigma_W^{<}(p_0,p_3,p_\bot|x) =  \int \frac{d^4 k}{(2\pi \hbar)^4}  i{\hbar^{-1}}g^2G^<_H(p-k)D^<(k) \nonumber\\    
				&\approx  i  \frac{w^2\hbar}{\rho_0^2}\int \frac{d^4 k}{(2\pi\hbar)^4} 
				e^{-\frac{(p_\bot-k_\bot - e A(x))^2}{|e\hbar B|}}2\pi \nonumber\\&\quad {\rm sign}\,(p_0-k_0) \delta((p_0-k_0)^2-(v_F(p_3-k_3))^2-m^2 v_F^4)\times\nonumber\\
				&\quad\times \Bigl((p_0-k_0)+v_F({p}_3 - {k}_3 )\alpha + \gamma^0 m v_F^2\Bigr)(1-i \gamma^1\gamma^2)\nonumber\\ &\quad \times  {n(p_0-k_0)} \,\pi\frac{\rho_0|\bk|}{u}\coth\frac{\beta u|\bk|}{2}\delta(k_0) \nonumber\\    
				&\approx  i  \frac{w^2\hbar}{\rho_0^2}\int \frac{d^3 k}{(2\pi\hbar)^4} 
				e^{-\frac{(p_\bot-k_\bot-eA(x))^2}{|e\hbar B|}}\nonumber\\
				&\quad 2\pi \delta(p_0^2-(v_F(p_3-k_3))^2-m^2v_F^4)\times\nonumber\\
				&\quad\times \Bigl(p_0+v_F({p}_3 - {k}_3 )\alpha^3 + \gamma^0 m v_F^2\Bigr)(1-i \gamma^1\gamma^2)\nonumber\\ &\quad {\rm sign}\,p_0\, {n(p_0)} \,\pi\frac{\rho_0|\bk|}{u}\coth\frac{\beta u|\bk|}{2} \nonumber\\
				&= i(\Sigma^{(0)}(p_0,p_3,p_\bot|x) + \Sigma^{(1)}_a(p_0,p_3,p_\bot|x)  \gamma^a \nonumber\\&+ \gamma^a\gamma^{b} \Sigma^{(2)}_{ab}(p_0,p_3,p_\bot|x) )2n(p_0) O^-
			\end{align}
			with the sum over $a,b = 0,1,2,3$ and
			\begin{align}
				&  \Sigma^{(0)}(p_0,p_3,p_\bot|x) ={\rm sign}\,p_0\, \frac{w^2}{\hbar^3\rho_0u} \frac{2 p_0}{8\pi^2\beta u}\nonumber\\
				&
				\int d^3k e^{-\frac{(p_\bot-k_\bot-eA(x))^2}{|e\hbar B|}}\nonumber\\
				&\delta(-p_0^2+({v_F}(p_3-k_3))^2+m^2 v_F^4)  \times t\coth t \Big|_{t=\frac{\beta u|\bk|}{2}} \nonumber\\
				&\approx \frac{w^2}{\hbar^3\rho_0u} \frac{1}{8\pi^2\beta u v_F}
				\int d^2k_\bot dk_3 e^{-\frac{(p_\bot-k_\bot-eA(x))^2}{|e\hbar B|}}\frac{|p_0|}{v_F|k_3-p_3|}\nonumber\\
				&\Bigl(\delta(\sqrt{p^2_0-m^2 v_F^4}/v_F - p_3 + k_3)\nonumber\\&+ \delta(-\sqrt{p^2_0-m^2 v_F^4}/v_F - p_3 + k_3)\Bigr)\Theta(p^2_0-m^2 v_F^4) \nonumber\\
				& \times (1+t^2/3)\Big|_{t=\frac{\beta u|\bk|}{2}}
			\end{align}
			\begin{align}
				&\approx \frac{w^2}{\hbar^3\rho_0u} \frac{1 }{ 4\pi^2  v_F}\frac{\Theta(p^2_0-m^2 v_F^4)}{\beta u \sqrt{1-\frac{m^2v_F^4}{p_0^2}}}
				\int d^2k_\bot  e^{-\frac{(p_\bot-k_\bot-eA(x))^2}{|e\hbar B|}} 
				\nonumber\\        
				&\approx  \frac{|e B| w^2 }{{4}\hbar^2\rho_0 \beta u^2} \frac{\Theta(p^2_0-m^2 v_F^4) }{ \pi  v_F \sqrt{1-\frac{m^2v_F^4}{p_0^2}}},
                \quad {\frac{u}{\fin{T}}\left(\frac{\hbar|e B|}{2\pi }\right)^{1/2} \ll 1}.
				\\    \Sigma^{(2)}_{03} &=\frac{w^2 {\rm sign}\,p_0\,}{\hbar^3\rho_0u} \frac{1}{8\pi^2}
				\int d^3k  e^{-\frac{(p_\bot-k_\bot-eA(x))^2}{|e \hbar B|}}\nonumber\\
				&\delta(p_0^2-{v_F}^2(p_3-k_3)^2-m^2 v_F^4) \times (p_3-k_3) |\bk|\coth\frac{\beta u|\bk|}{2} \nonumber\\
				&\approx 0 , \quad \frac uT\left(\frac{|e \hbar B|}{2\pi }\right)^{1/2} \ll 1 \nonumber\\
				\Sigma^{(1)}_0 &=\frac{w^2 {\rm sign}\,p_0\,}{\hbar^3\rho_0u} \frac{mv_F^2}{{8}\pi^2}
				\int d^3k e^{-\frac{(p_\bot-k_\bot-eA(x))^2}{|e \hbar B|}}\nonumber\\
				&\delta(p_0^2-v^2_F(p_3-k_3)^2-m^2v_F^4) \times|\bk|\coth\frac{\beta u|\bk|}{2} \nonumber\\ 
				&\approx  \frac{|e B| w^2 m v_F^2 }{{4}\hbar^2\rho_0 p_0\beta u^2} \frac{\Theta(p^2_0-m^2 v_F^4) }{ \pi  v_F \sqrt{1-\frac{m^2v_F^4}{p_0^2}}}
		\end{align}}
		The other components of $\Sigma^{(1,2)}$ are zero. 
        {The given approximation may be justified at sufficiently large temperatures: 
			\begin{equation}
                \kappa = \frac{u}{\fin{T}}\left(\frac{|e \hbar B|}{2\pi }\right)^{1/2} = \frac{1}{\sqrt{2\pi}}\frac{u}{v_F}\frac{l_T}{l_B} \label{kappa}
			\end{equation}
            For example, {for  ZrTe$_5$}, Fermi velocity $v_F \sim 10^6$ m/s, $u \sim 10^3$ m/s, and at $T = 30 $K, and $B = 1 $T we have $l_B \sim 6 \times 10^{-8}m$, $l_T\sim  10^{-5}m$, which gives 
{    
\begin{equation}
    \kappa \sim 0.16, \label{EqKappaZrTe5}
\end{equation}
}
while at $T=10$K we would have $\kappa \sim 0.5$.  }

		We arrive at the following expression for the lesser component of Keldysh self-energy in the one - loop approximation
		{\begin{eqnarray}
				\Sigma^< &=& \chi \mu  \Bigl(1  + \frac{mv_F^2}{p_0}\gamma^0\Bigr) \frac{2 i n(p_0)}{\sqrt{1-\frac{m^2v_F^4}{p_0^2}}}  \Theta(p^2_0-m^2 v_F^4)
		\end{eqnarray}}
		We assume that parameter $\chi(B,T,\mu) = \frac{|e B| w^2 T}{ {4}\pi \hbar^2 u^2  v_F\rho_0 \mu} = \frac{w^2}{{4}\pi l_B^2 l_T u^2 \rho_0 \mu} \ll 1$ remains small, which results in smallness of the given correction. {For $ZrTe_5$ with $\rho_0 \sim 5$ g/cm we obtain (estimating roughly $w \sim 2$ eV) at $T=50$K, $B=10$T, $\mu \sim 100 $meV: $\chi \sim 2\times 10^{-3}$.}
			At {$\mu^2 \gg m^2v_F^4$} we may effectively substitute to the above expression $p_0 \approx v_F|p_3| \approx \mu$ and neglect the second term:
			\begin{eqnarray}
				{\Sigma}^< &=&  2 i\, \chi  \mu    n(p_0)  , 
			\end{eqnarray}
			{which is equivalent to a }small dissipation rate 
			\begin{equation}
                \epsilon^{{\text{ph}}}_B \approx \frac{|e B| w^2 T}{{4}\pi \hbar^2 u^2  v_F\rho_0 }\label{epsilonB}.
			\end{equation}

\subsection{Calculation of full self-energy}\label{SectAppC}
			
Here we calculate the contribution of phonons to the component $\Sigma^R = i{\hbar^{-1}}g^2(G^<D^R+G^RD^<+G^RD^R)$ of self-energy
    (the other components can be restored from general analytical properties of the first order self-energy correction, see \cite{LL10}). 
First, we calculate the contribution similar to that of the lesser component $\Sigma^{R}_I = i{\hbar^{-1}}g^2(G^RD^<)$:
\begin{align}
    & \Sigma_{I,W}^{R}(p_0,p_3,p_\bot|x) =  \int \frac{d^4 k}{(2\pi \hbar)^4}  i{\hbar^{-1}}g^2G^R_W(p-k)D^<(k) \nonumber\\    
    &\approx    \frac{w^2\hbar}{\rho_0^2}\int \frac{d^4 k}{(2\pi\hbar)^4} 
        e^{-\frac{(p_\bot-k_\bot - e A(x))^2}{|e\hbar B|}} \nonumber\\&\quad  ((p_0-k_0+i\epsilon)^2-(v_F(p_3-k_3))^2-m^2 v_F^4)^{-1}\times\nonumber\\
	&\quad\times \Bigl((p_0-k_0)+v_F({p}_3 - {k}_3 )\alpha + \gamma^0 m v_F^2\Bigr)(1-i \gamma^1\gamma^2)\nonumber\\ &\quad \times  \,\pi\frac{\rho_0|\bk|}{u}\coth\frac{\beta u|\bk|}{2}\delta(k_0) \nonumber\\    
	&\approx    \frac{w^2\hbar}{\rho_0^2}\int \frac{d^3 k}{(2\pi\hbar)^4} 
	    e^{-\frac{(p_\bot-k_\bot-eA(x))^2}{|e\hbar B|}} \times\nonumber\\
	&\quad\times \frac{\Bigl(p_0+v_F({p}_3 - {k}_3 )\alpha^3 + \gamma^0 m v_F^2\Bigr)}{((p_0+i\epsilon)^2-(v_F(p_3-k_3))^2-m^2v_F^4)}(1-i \gamma^1\gamma^2)\nonumber\\ &\quad   \,\pi\frac{\rho_0|\bk|}{u}\coth\frac{\beta u|\bk|}{2} \nonumber\\
	&= i(\Sigma^{R(0)}_I(p_0,p_3,p_\bot|x) + \Sigma^{R(1)}_{I,a}(p_0,p_3,p_\bot|x)  \gamma^a \nonumber\\&+ \gamma^a\gamma^{b} \Sigma^{R(2)}_{I,ab}(p_0,p_3,p_\bot|x) ) O^-
\end{align}
with the sum over $a,b = 0,1,2,3$ and
\begin{align}
    \Sigma^{R(0)}_I&(p_0,p_3,p_\bot|x) = \frac{iw^2}{\pi \hbar^3\rho_0u} \frac{ p_0}{4\pi^2\beta u}\int d^3k \times\nonumber\\
	&\times (-(p_0+i\epsilon)^2+({v_F}(p_3-k_3))^2+m^2 v_F^4)^{-1}\times\nonumber\\  
    &\times e^{-\frac{(p_\bot-k_\bot-eA(x))^2}{|e\hbar B|}}
        ~t\coth t \Big|_{t=\frac{\beta u|\bk|}{2}} \nonumber \\
	&\approx -\frac{w^2}{\hbar^3\rho_0u} \frac{1 }{ 4\pi^2  v_F}\frac{1}{\beta u \sqrt{1-\frac{m^2v_F^4}{p_0^2}}} \times\nonumber\\
    &\qquad\times\int d^2k_\bot  e^{-\frac{(p_\bot-k_\bot-eA(x))^2}{|e\hbar B|}} 
	\nonumber\\        
	&\approx  -\frac{|e B| w^2 }{4\hbar^2\rho_0 \beta u^2} \frac{1 }{ \pi  v_F \sqrt{1-\frac{m^2v_F^4}{p_0^2}}},
\end{align}
at {\(\kappa = \frac{u}{\fin{T}}\left(\frac{\hbar|e B|}{2\pi }\right)^{1/2} \ll 1\), see above discussion after Eq. (\ref{kappa})}.
\begin{align}
	\Sigma^{(2)}_{03} &=-i\frac{w^2 {\rm sign}\,p_0\,}{\pi\hbar^3\rho_0u} \frac{1}{4\pi^2}
	\int d^3k  e^{-\frac{(p_\bot-k_\bot-eA(x))^2}{|e\hbar B|}}\nonumber\\
	&\frac{ (p_3-k_3) |\bk|\coth\frac{\beta u|\bk|}{2}}{(p_0+i\epsilon)^2-{v_F}^2(p_3-k_3)^2-m^2 v_F^4} \nonumber\\
	&\approx 0 , \quad \frac{u}{\fin{T}}\left(\frac{|e\hbar B|}{2\pi }\right)^{1/2} \ll 1 \nonumber\\
	\Sigma^{R(1)}_{I,0} &=-i \frac{w^2 }{\pi\hbar^3\rho_0u} \frac{mv_F^2}{4\pi^2}
	\int d^3k e^{-\frac{(p_\bot-k_\bot-eA(x))^2}{|e\hbar B|}}\nonumber\\
	&\frac{|\bk|\coth\frac{\beta u|\bk|}{2}}{((p_0+i\epsilon)^2-{v_F}^2(p_3-k_3)^2-m^2 v_F^4} \nonumber\\ 
	&\approx  -\frac{|e B| w^2 m v_F^2 }{4\hbar^2\rho_0 p_0\beta u^2} \frac{1 }{ \pi  v_F \sqrt{1-\frac{m^2v_F^4}{p_0^2}}}
\end{align}
We arrive at 
\begin{eqnarray}
    {\Sigma}^{R}_I &=& -\chi \mu  \Bigl(1  + \frac{mv_F^2}{p_0}\gamma^0\Bigr) \frac{1}{\sqrt{1-\frac{m^2v_F^4}{p_0^2}}} i O^-
\end{eqnarray}
			
For the second part $\Sigma^{R}_{II} = i{\hbar^{-1}}g^2(G^RD^R)$ we obtain
\begin{align}
	& \Sigma_{II,W}^{R}(p_0,p_3,p_\bot|x) =  \int \frac{d^4 k}{(2\pi \hbar)^4}  i{\hbar^{-1}}g^2G^R_W(p-k)D^R(k) \nonumber\\    
	&\approx    i\frac{w^2\hbar}{\rho_0^2}\int \frac{d^4 k}{(2\pi\hbar)^4} 
	e^{-\frac{(p_\bot-k_\bot - e A(x))^2}{|e\hbar B|}} \nonumber\\&\quad  ((p_0-k_0+i\epsilon)^2-(v_F(p_3-k_3))^2-m^2 v_F^4)^{-1}\times\nonumber\\
	&\quad\times \Bigl((p_0-k_0)+v_F({p}_3 - {k}_3 )\alpha + \gamma^0 m v_F^2\Bigr)(1-i \gamma^1\gamma^2)\nonumber\\ &\quad \times \frac{\rho_0k^2}{(k_0+i\epsilon)^2-u^2k^2}  \nonumber\\    
	&\approx    -\frac{w^2}{\rho_0^2}\int \frac{d^2 k_\bot dk_3}{(2\pi\hbar)^3} 
	e^{-\frac{(p_\bot-k_\bot-eA(x))^2}{|e\hbar B|}} O^-\times\nonumber\\
	\times& \Bigl[ \frac{\Bigl(-\sqrt{v_F^2(p_3 - k_3)^2+m^2v_F^4}+v_F({p}_3 - {k}_3 )\alpha^3 + \gamma^0 m v_F^2\Bigr)}{\sqrt{v_F^2(p_3 - k_3)^2+m^2v_F^4}}\nonumber\\ &\quad   \,\frac{\rho_0k^2}{(p_0 + \sqrt{v_F^2(p_3 - k_3)^2+m^2v_F^4}+i\epsilon)^2 - u^2 k^2} \nonumber\\
	-& \frac{\Bigl(\sqrt{v_F^2(p_3 - k_3)^2+m^2v_F^4}+v_F({p}_3 - {k}_3 )\alpha^3 + \gamma^0 m v_F^2\Bigr)}{\sqrt{v_F^2(p_3 - k_3)^2+m^2v_F^4}}\nonumber\\ &\quad   \,\frac{\rho_0k^2}{(p_0 - \sqrt{v_F^2(p_3 - k_3)^2+m^2v_F^4}+i\epsilon)^2 - u^2 k^2}\Bigr] 
\end{align}
For the third part $\Sigma^{R}_{III} = i{\hbar^{-1}}g^2(G^<D^R)$ we obtain
\begin{align}
	& \Sigma_{III,W}^{R}(p_0,p_3,p_\bot|x) =  \int \frac{d^4 k}{(2\pi \hbar)^4}  i{\hbar^{-1}}g^2G^<_W(p-k)D^R(k)  \nonumber\\    
	&\approx    -2\pi \frac{w^2\hbar}{\rho_0^2}\int \frac{d^4 k}{(2\pi\hbar)^4} 
	e^{-\frac{(p_\bot-k_\bot - e A(x))^2}{|e\hbar B|}}{\rm sign}\,(p_0-k_0) \nonumber\\&\quad  \delta((p_0-k_0)^2-(v_F(p_3-k_3))^2-m^2 v_F^4)\times\nonumber\\
	&\quad\times n(p_0-k_0) \Bigl((p_0-k_0)+v_F({p}_3 - {k}_3 )\alpha + \gamma^0 m v_F^2\Bigr)\nonumber\\ 
    &\quad \times (1-i \gamma^1\gamma^2)\frac{\rho_0k^2}{(k_0+i\epsilon)^2-u^2k^2}  \nonumber\\    
	&\approx    -\frac{w^2}{\rho_0^2}\int \frac{d^2 k_\bot dk_3}{(2\pi\hbar)^3} 
	e^{-\frac{(p_\bot-k_\bot-eA(x))^2}{|e\hbar B|}} O^-\times\nonumber\\
	\times& \Bigl( -\frac{-\sqrt{v_F^2(p_3 - k_3)^2+m^2v_F^4}+v_F({p}_3 - {k}_3 )\alpha^3 + \gamma^0 m v_F^2}{\sqrt{v_F^2(p_3 - k_3)^2+m^2v_F^4}}\nonumber\\ & \quad \frac{\rho_0 k^2 n(-\sqrt{v_F^2(p_3 - k_3)^2+m^2v_F^4})}{(p_0 + \sqrt{v_F^2(p_3 - k_3)^2+m^2v_F^4}+i\epsilon)^2 - u^2 k^2}\nonumber\\&
	+\frac{(\sqrt{v_F^2(p_3 - k_3)^2+m^2v_F^4}+v_F({p}_3 - {k}_3 )\alpha^3 + \gamma^0 m v_F^2}{\sqrt{v_F^2(p_3 - k_3)^2+m^2v_F^4}}\nonumber\\ 
    & \frac{\rho_0k^2 n(\sqrt{v_F^2(p_3 - k_3)^2+m^2v_F^4})}{(p_0 - \sqrt{v_F^2(p_3 - k_3)^2+m^2v_F^4}+i\epsilon)^2 - u^2 k^2}\Bigr) 
\end{align}
We combine the two contributions and obtain for their sum
\begin{eqnarray}
	\Sigma^{R}_{II} + \Sigma^{R}_{III}	&=& (\Sigma^{R(0)}_{II,III}(p_0,p_3,p_\bot|x) +\\
    &&+ \Sigma^{R(1)}_{II,III,a}(p_0,p_3,p_\bot|x)  \gamma^a \nonumber\\
    &&+ \gamma^a\gamma^{b} \Sigma^{R(2)}_{II,III,ab}(p_0,p_3,p_\bot|x) ) O^-,\nonumber
\end{eqnarray}
with the sum over $a,b = 0,1,2,3$ and on the Fermi surface (at $T\to 0,\mu \gg mv_F^2$)
\begin{align}
	&  \Sigma^{R(0)}_{II,III}(p_0,p_3,p_\bot|x) = \frac{w^2}{(2\pi \hbar)^3\rho_0} 
	\int k^2d^3k e^{-\frac{(p_\bot-k_\bot-eA(x))^2}{|e\hbar B|}}\nonumber\\
	&\Bigl[\frac{1- n(-\sqrt{v_F^2(p_3 - k_3)^2+m^2v_F^4})}{(p_0 + \sqrt{v_F^2(p_3 - k_3)^2+m^2v_F^4}+i\epsilon)^2 - u^2 k^2}\nonumber\\& 
	\frac{1- n(\sqrt{v_F^2(p_3 - k_3)^2+m^2v_F^4})}{(p_0 - \sqrt{v_F^2(p_3 - k_3)^2+m^2v_F^4}+i\epsilon)^2 - u^2 k^2}\Bigr]\nonumber\\&
	= \frac{w^2}{(2\pi \hbar)^3\rho_0} 
	\int (k_3^2 + k_\bot^2)dk_3 d^2k_\bot e^{-\frac{(p_\bot-k_\bot-eA(x))^2}{|e\hbar B|}}\nonumber\\
	\times& 
	\frac{1- n(\sqrt{v_F^2(p_3 - k_3)^2+m^2v_F^4})}{(p_0 - \sqrt{v_F^2(p_3 - k_3)^2+m^2v_F^4}+i\epsilon)^2 - u^2 (k_3^2+k_\bot^2)}
\end{align}
Several comments are in order concerning this expression. 

{First of all, at $T{\ll}\sqrt{v_F^2(p_3 - k_3)^2+m^2v_F^4}<|\mu|$ the expression standing inside the integral vanishes. Only those  electrons contribute the self-energy, for which the energy is larger than the Fermi energy. This means that the  electron passes from the Fermi surface to one of the vacant places with ejection of virtual phonon, then it returns back to the Fermi surface with absorption of the same virtual phonon.}
			
Next, the integral over $k_3$ is divergent in ultraviolet. Therefore, we should introduce the new parameter of the model - the phonon energy $\Lambda$, at which our approximation for the form of phonon propagator ceases to work (very roughly $\Lambda$ may be estimated is an energy of optical phonon). The corresponding length scale is $l_\Lambda = \frac{\hbar u}{\Lambda}$. 
			
These notices allow us to make an estimate:
\begin{align}
    \Sigma^{R(0)}_{II,III}(p_0,p_3,p_\bot|x) &\approx \frac{2 \pi w^2 \Lambda \hbar |e B|}{(2\pi \hbar)^3\rho_0 v_F^2 u} \nonumber\\ &= \frac{w^2 }{4 \pi^2 l_B^2 l_\Lambda \rho_0 v_F^2 } 
\end{align}
This estimate should be compared to $\Sigma^{R(0)}_{I}\sim \chi \mu \sim \frac{w^2}{4 \pi l_B^2 l_T u^2 \rho_0 } $. 
One can see that 
\begin{equation}
    \Sigma^{R(0)}_{II,III}/\Sigma^{R(0)}_{I} \sim \frac{1}{\pi} \frac{l_T}{l_\Lambda}\frac{u^2}{v^2_F}
\end{equation}
{Again, we use for the reference   Cd$_3$As$_2$ with Fermi velocity $v_F \sim 10^6$ m/s, $u \sim 10^3$ m/s,  at $T = 30 $K we have  $l_T\sim 0.3\times 10^{-5}m$. Estimating roughly $\Lambda \sim 5 $ meV, we get $l_\Lambda \sim 5\times 10^{-10}$m,  and $\Sigma^{R(0)}_{II,III}/\Sigma^{R(0)}_{I}\sim {10^{-2}}$.} Therefore, we neglect this contribution in our estimates. 
			
The direct calculation shows that $\Sigma^{R (03)}_{II,III}$ vanishes identically. For $\Sigma^{R (1)}_{II,III}$ we obtain:
\begin{align}
	&  \Sigma^{R(1)}_{II,III}(p_0,p_3,p_\bot|x) = \frac{w^2}{(2\pi \hbar)^3\rho_0} 
	\int k^2d^3k e^{-\frac{(p_\bot-k_\bot-eA(x))^2}{|e\hbar B|}}\nonumber\\
	&\Bigl[\frac{-1+ n(-\sqrt{v_F^2(p_3 - k_3)^2+m^2v_F^4})}{(p_0 + \sqrt{v_F^2(p_3 - k_3)^2+m^2v_F^4}+i\epsilon)^2 - u^2 k^2}\nonumber\\& 
	\frac{1- n(\sqrt{v_F^2(p_3 - k_3)^2+m^2v_F^4})}{(p_0 - \sqrt{v_F^2(p_3 - k_3)^2+m^2v_F^4}+i\epsilon)^2 - u^2 k^2}\Bigr]\nonumber\\& \frac{m v_F^2}{\sqrt{v_F^2(p_3 - k_3)^2+m^2v_F^4})}\nonumber\\&
	= \frac{w^2}{(2\pi \hbar)^3\rho_0} 
	\int (k_3^2 + k_\bot^2)dk_3 d^2k_\bot e^{-\frac{(p_\bot-k_\bot-eA(x))^2}{|e\hbar B|}}\nonumber\\
	\times& \frac{1- n(\sqrt{v_F^2(p_3 - k_3)^2+m^2v_F^4})}{(p_0 - \sqrt{v_F^2(p_3 - k_3)^2+m^2v_F^4}+i\epsilon)^2 - u^2 (k_3^2+k_\bot^2)}
	\nonumber\\&\frac{m v_F^2}{\sqrt{v_F^2(p_3 - k_3)^2+m^2v_F^4})}
\end{align}
Here the integral over $k_3$ is logarithmically divergent, and we arrive at the estimate 
\begin{align}
	\Sigma^{R(1)}_{II,III}(p_0,p_3,p_\bot|x) &\approx \frac{2 \pi w^2 m v_F^2 \hbar |e B|}{(2\pi \hbar)^3\rho_0 v_F^3}\, {\rm ln}\,\Bigl(\frac{\Lambda}{mv^2_F}\frac{v_F}{u}\Bigr) \nonumber\\ 
= \frac{w^2 }{4 \pi^2 l_B^2 \rho_0 v_F^2 }&\,\frac{m v_F}{\hbar} \, {\rm ln}\,\Bigl(\frac{\Lambda}{mv^2_F}\frac{v_F}{u}\Bigr)
\end{align}
One can see that 
\begin{equation}
	\Sigma^{R(1)}_{II,III}/\Sigma^{R(1)}_{I} \sim \frac{1}{\pi} \frac{l_T }{l_\mu }\frac{u^2}{v^2_F}\,{\rm ln} \,\Bigl(\frac{\Lambda}{mv^2_F}\frac{v_F}{u}\Bigr)
\end{equation}
and 
\begin{equation}
	\Sigma^{R(1)}_{II,III}/\Sigma^{R(0)}_{I} \sim \frac{1}{\pi} \frac{l_T mv_F }{\hbar }\frac{u^2}{v^2_F}\,{\rm ln} \,\Bigl(\frac{\Lambda}{mv^2_F}\frac{v_F}{u}\Bigr)
\end{equation}
{This ratio may be even smaller than $\Sigma^{R(0)}_{II,III}/\Sigma^{R(0)}_{I}$ provided that mass parameter $m$ is small compared to $\mu$ as assumed in the present paper.} 

We arrive finally at 
\begin{eqnarray}
	{\Sigma}^{R} &\approx& -i\chi \mu  \Bigl(1  + \frac{mv_F^2}{p_0}\gamma^0\Bigr) \frac{1}{\sqrt{1-\frac{m^2v_F^4}{p_0^2}}}.
\end{eqnarray}

\bibliography{CMEwKeldysh_ext,biblio_corrected}

\end{document}